\documentclass[bibyear]{aa}  

\usepackage{lipsum}
\usepackage{subcaption}         
\usepackage{lscape}             
\usepackage{placeins}           
\usepackage[toc]{appendix} 
\usepackage{txfonts}
\usepackage{graphicx}	
\usepackage{amsmath}	
\usepackage{mathrsfs}
\usepackage[breaklinks, colorlinks, citecolor=blue, urlcolor = blue]{hyperref}
\usepackage{pdflscape}
\usepackage{float}
\usepackage{here}
\usepackage[export]{adjustbox}
\usepackage{array}
\usepackage{natbib}
\usepackage{orcidlink}
\bibpunct{(}{)}{;}{a}{}{,} 

\begin{document}

   \title{Stellar bars reshape the spatial distribution of UV-bright clumps in nearby galaxies}
   \titlerunning{Bars and UV clump distributions}
   \authorrunning{Zhou et al.}

%
%
%
    \author{Zhimin Zhou \inst{1} \orcidlink{0000-0002-4135-0977} \corrauth{zmzhou@nao.cas.cn}
	\and Jianfei Liu\inst{1,2}\email{liujf@bao.ac.cn}
    }

    \institute{National Astronomical Observatories, Chinese Academy of Sciences, Beijing 100101, People’s Republic of China
    \and School of Astronomy and Space Science, University of Chinese Academy of Sciences, Beijing 100049, People’s Republic of China
    }
    
   \date{Received XXX; accepted YYY}

    \abstract
    {Stellar bars are important drivers of secular evolution in disk galaxies. By redistributing gas and angular momentum, they can influence recent star formation, but their dominant effect on star-forming regions remains uncertain.}
    {We investigated how stellar bars affect the radial and azimuthal distributions of star-forming clumps and whether they modify the observed luminosities and sizes of individual clumps. We also examined how these effects depend on host-galaxy stellar mass and bar properties.}
    {We analyzed archival HST/WFC3 F275W imaging of 70 barred galaxies and 95 unbarred control galaxies matched in stellar mass, optical color, redshift, and image quality. Using a uniform dendrogram-based method, we identified 95\,486 star-forming clumps bright in the ultraviolet. We compared their radial and azimuthal distributions, central F275W emission, and luminosity and size distribution functions.}
    {Barred galaxies show a clear deficit of UV-bright clumps at intermediate radii broadly corresponding to the bar region and a local recovery toward the bar ends. Their clumps are also preferentially concentrated along the bar major axis. These spatial differences are stronger in high-mass galaxies. The normalized bar length shows a clearer association with the radial redistribution and central F275W concentration than bar ellipticity, although its relation with the azimuthal distribution is less uniform. When the radius is normalized by bar size, the local recovery occurs near $R/R_{\rm bar}\sim1$, suggesting that the bar provides a characteristic spatial scale for the redistribution. Despite these spatial differences, barred and control galaxies have broadly similar clump luminosity and size distribution functions.}
    {Stellar bars primarily regulate the spatial distribution of recent star formation on galactic scales rather than substantially changing the observed luminosities and sizes of individual UV clumps. In addition, the strength of this redistribution depends on host-galaxy mass and bar extent.}

    \maketitle
    \nolinenumbers

\section{Introduction} 
\label{sec:intro}
Stellar bars are common non-axisymmetric structures in disk galaxies. In the local Universe, reported bar fractions in disk galaxies range from approximately 30\% to 70\%, depending on sample selection, observing wavelength, and classification method \citep{Nair2010, Masters2011, Erwin2018, Wang2025}. At higher redshifts, this bar fraction declines steadily, yet recent space-based observations confirm its persistence out to $z \sim 4$ \citep{Melvin2014, Guo2025, Geron2025, LeConte2026}. Their high incidence and long lifetimes make stellar bars an important driver of the secular evolution of galaxies across cosmic time \citep[][and references therein]{Kormendy2004}.

Bars modify galaxies by redistributing angular momentum through non-axisymmetric perturbations in the gravitational potential \citep{Regan2004, Athanassoula2014}. Hydrodynamical simulations and observations have shown that bar-driven gravitational torques can efficiently transport gas from the outer disk into the central regions \citep{Sheth2005, Athanassoula2013, Kobayashi2026}. This inward migration can trigger central star formation \citep{Wang2012, Verwilghen2024, Lokas2025}, and promote the formation of nuclear rings, pseudobulges, and central mass concentrations \citep[e.g.,][]{Ellison2011, Lin2020, Romeo2023, Kwak2026}. Furthermore, the sustained gas inflow can channel material toward the central supermassive black hole, potentially fueling active galactic nuclei \citep[e.g.,][]{Ho1997, Garcia-Burillo2005, Verwilghen2024, Kataria2024, Liu2026}. Consequently, the presence of a bar tightly links large-scale orbital dynamics to the small-scale evolution of baryonic matter.

Because bars redistribute cold gas, they are expected to affect both the intensity and the spatial distribution of star formation. Yet the relation between stellar bars and star formation remains unsettled. Barred galaxies have indeed been found to show enhanced star formation in their central regions, along the bar, or in circumnuclear rings, consistent with gas inflow driven by the non-axisymmetric bar potential \citep{Coelho2011, Zhou2015, Chown2019, Zee2023}. 
Conversely, other systematic studies have reported that barred and unbarred galaxies have similar global star-forming properties \citep[e.g.,][]{Aguerri2009, Willett2015}.
Other studies have instead reported reduced star formation within the bar region or lower global activity, especially in massive, red, or gas-poor systems \citep[e.g.,][]{Khoperskov2018, Fraser-McKelvie2020a, Newnham2020, Kim2024a, Scaloni2024}.
These seemingly conflicting results suggest that the influence of stellar bars on star formation is not universal but likely depends on both the physical properties of the host galaxy and the evolutionary state of the bar itself \citep{Oh2012, Masters2012, Diaz-Garcia2021, Geron2021}.

A further limitation in understanding the connection between stellar bars and star formation is the widespread use of integrated or centrally averaged measurements. Although these quantities provide valuable constraints on the overall level of star formation, they offer only limited information on where recent star formation occurs within galaxies. Consequently, they cannot distinguish whether bars primarily enhance or suppress star formation or instead redistribute star-forming activity across galactic disks. Addressing this question requires spatially resolved tracers of recent star formation \citep[e.g.,][]{Neumann2019, Diaz-Garcia2020, Renu2025}.

In disk galaxies, star formation is inherently clustered and occurs within discrete star-forming complexes. These complexes can be readily identified in deep high-resolution ultraviolet images, where the emission is dominated by short-lived OB stars with lifetimes of less than $\sim$ 100 Myr \citep{PortegiesZwart2010, Guo2015, Shashank2026}. Compared with optical tracers, UV emission is much less affected by contamination from older stellar populations \citep{Hao2011, Kennicutt2012}. Statistical analyses of individual star-forming clumps therefore offer a powerful way to investigate how stellar bars influence the spatial distribution of recent star formation on sub-galactic scales.

Despite their potential, studies directly connecting stellar bars with individual star-forming clumps remain scarce, primarily due to the spatial resolution limits of wide-field surveys. However, the high-resolution F275W imaging from the Hubble Space Telescope (HST) provides the angular resolution needed to identify individual UV-bright clumps. Operating at near-UV (NUV) wavelengths, this filter probes physical scales down to a few hundred parsecs or better in nearby galaxies \citep{Calzetti2015, Lee2022, Thilker2025}. These capabilities enable detailed investigation of young stellar populations. Furthermore, such HST observations allow for examination of how localized star formation responds to the presence and properties of stellar bars.

In this work, we investigate how stellar bars influence the spatial distribution of recent star formation using high-resolution HST F275W imaging. Our sample consists of nearby barred galaxies and a matched control sample of unbarred systems. We identified individual star-forming clumps in each galaxy to compare their radial and azimuthal distributions. These profiles were then further analyzed as a function of galaxy stellar mass, normalized bar length, and bar ellipticity. Additionally, we evaluated whether the presence of a bar shifts the luminosity functions and size distributions of these star-forming clumps. These analyses allowed us to determine whether stellar bars alter the observed properties of star-forming clumps or mainly reshape their spatial distribution within galaxies.

The paper is organized as follows. In Section \ref{sec:data} we describe the sample selection, data, and bar measurements. Section \ref{sec:method} presents the clump identification and photometry procedures. Our main results are presented in Section \ref{sec:analysis} and followed by a detailed physical discussion in Section \ref{sec:discuss}. Finally, in Section \ref{sec:summary} we summarize our primary conclusions. Throughout this paper, we adopt a standard Lambda cold dark matter ($\Lambda$CDM) cosmology with $\it {H}_{\rm 0} {\rm= 70\ km\ s^{-1}\ Mpc^{-1}}$, $\rm {\Omega_M = 0.3}$, and $\rm{\Omega_{\Lambda} = 0.7}$.

\section{Sample and data}
\label{sec:data}

\subsection{Parent sample and HST imaging}
Our parent sample was constructed using archival data from the Mikulski Archive for Space Telescopes (MAST) $\footnote{\url{https://mast.stsci.edu/search/ui/##/hst}}$. We searched for nearby galaxies observed with the HST/WFC3 UVIS F275W filter. This wide-band NUV filter is centered at approximately $2710\text{ \AA}$, making it well suited to tracing young massive stars and recent star formation.

We initially collected $\sim$ 300 galaxies within the redshift range of $0.01 < z < 0.1$. This redshift interval provides a suitable balance between spatial resolution and field coverage. At these redshifts, the angular resolution of HST is sufficient to resolve individual star-forming clumps, while the lower redshift limit minimizes the likelihood of resolving individual stars rather than star-forming complexes. Interacting, merging, highly disturbed and edge-on galaxies were excluded based on visual inspection of their optical morphologies to facilitate reliable identification of bars and clumps. This selection also helps reduce confusion between bar-related and interaction-induced star formation, although this may also exclude galaxies whose bars were affected by interactions. Thus, our conclusions apply to galaxies without obvious interaction signatures, although weak or past interactions cannot be ruled out. The final parent sample consists of 245 galaxies.

The archival observations were obtained primarily from HST programs 14779, 16287, 16657, 16741, and 17179. The images were processed with the standard HST pipeline and combined using AstroDrizzle. Throughout this work, we use the \texttt{\_drc} products, which are flat-fielded, geometrically corrected, and corrected for charge transfer efficiency (CTE). The final frames have a uniform pixel scale of $0\farcs04$ and a stable point-spread function (PSF) with an a full width at half maximum (FWHM) of $0\farcs07$, resolving physical scales of $\sim$ 20--120 pc across the redshifts of our sample.

\subsection{Optical imaging, galaxy properties, and bar measurements}
\label{sec:Optical}
Complementary optical imaging was obtained to characterize the stellar structure and global properties of the parent sample. We primarily used the $g, r$, and $z$-band images from the Dark Energy Spectroscopic Instrument (DESI) Legacy Imaging Surveys Data Release 10 (DR10; \citealt{Dey2019, Moustakas2023}). For the small number of galaxies outside the DESI footprint, equivalent $grizy$ images from the Pan-STARRS1 survey (PS1; \citealt{Chambers2016}) were used instead.

Before the structural analysis, foreground stars and unrelated background sources were masked following the empirical procedure described in \citet{Zhou2020}. 
Integrated galaxy photometry was then measured from the masked images. All optical fluxes were corrected for Galactic foreground extinction using the dust maps of \citet{Schlafly2011} and a \citet{Fitzpatrick1999} extinction curve with $R_V = 3.1$. Total stellar masses ($M_{\star}$) were estimated from the extinction-corrected $g-r$ color and $r$-band luminosity using the color-dependent mass-to-light ratio calibration of \citet{Bell2003}.

Stellar bars were identified following the ellipse-fitting procedure described in \citet{Wang2025}. In brief, barred galaxies were selected based on the characteristic radial behavior of ellipticity and position angle profiles. The bar length ($R_{\rm bar}$) was taken as the semi-major axis corresponding to the maximum ellipticity, and the bar ellipticity ($e_{\rm bar}$) was measured at the same radius. 

Applying this procedure, we identified 82 barred galaxies and measured their structural parameters.  Both $R_{\rm bar}$ and $e_{\rm bar}$ were corrected for projection effects. Throughout this work, the normalized bar length is defined as $R_{\rm bar}/R_{\rm eff}$, where $R_{\rm eff}$ is the optical effective radius of the host galaxy.
The remaining parent-sample galaxies comprise 132 classified as unbarred and 31 with uncertain bar classifications. The uncertain cases were excluded from the barred--control analyses. The 82 identified barred galaxies account for 33.5\% of the full parent sample. This fraction reflects our archival selection and classification criteria rather than an unbiased estimate for nearby disk galaxies.

For comparison with a broader galaxy sample, we selected disk and barred galaxies from SDSS/Galaxy Zoo 2 (GZ2) using morphological vote fractions. Over the same redshift range, our barred sample overlaps substantially with the GZ2 in stellar mass and optical color. However, our galaxies tend to have lower stellar masses and bluer $g-r$ colors. These differences in sample coverage are relevant to the interpretation of the stellar-mass trends discussed below.

\begin{figure}
    \centering
    \includegraphics[width=\columnwidth]{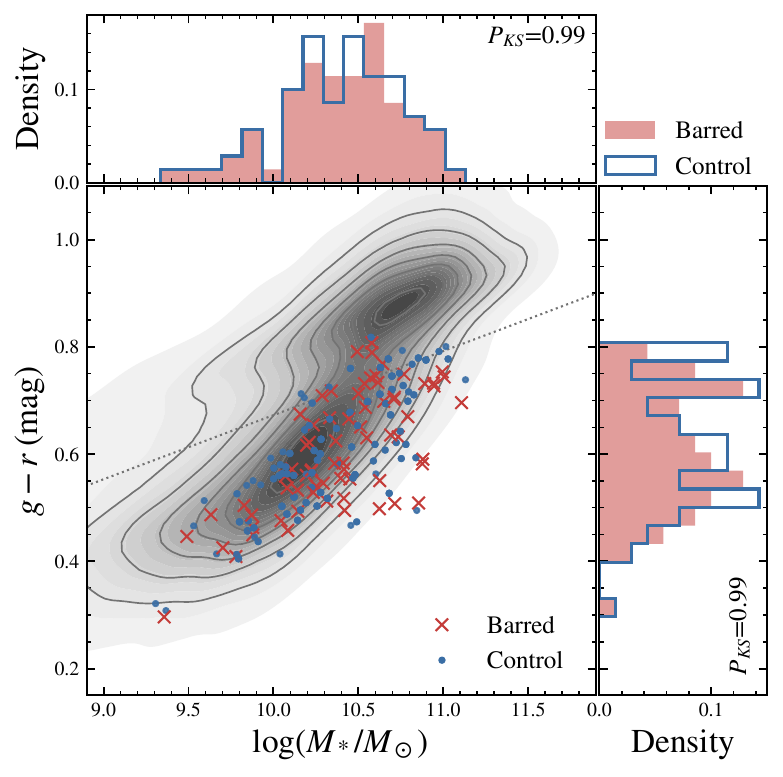}
	\caption{Distribution of the matched barred and control galaxies in the stellar mass–color ($M_{\star}$–$g-r$) plane. The gray background shows the galaxy distribution from the SDSS DR7 MPA/JHU catalog \citep{Kauffmann2003, Salim2007}, with number-density contours overlaid to illustrate the underlying galaxy population. The dashed line shows the demarcation between passive and star-forming galaxies.
    The upper and right panels present the corresponding distributions of stellar mass and $g-r$ color for the barred and control samples. The $p$-values from two-sample KS tests are indicated in the respective panels. The high KS test $p$-values indicate that the matched barred and control samples are statistically consistent in both stellar mass and optical color.
	\label{fig:matched_mass_color}}
\end{figure}

\subsection{Matched control sample}
\label{sec:matched_control}
To isolate the impact of stellar bars on recent star formation, we constructed a control sample from the 132 parent-sample galaxies classified as unbarred. For the spatial-profile analysis, each barred galaxy was matched to unbarred candidates that satisfied stellar mass $\Delta \log(M_{\star}) \le 0.1\text{ dex}$, color $\Delta(g-r) \le 0.05\text{ mag}$, and redshift $\Delta(z) \le 0.05$. 
We additionally matched the F275W image depth using the pixel-to-pixel background dispersion, $\sigma_{\rm sky}$, thereby limiting differences in image depth and clump-detection sensitivity between the barred and control samples. Candidate controls were required to satisfy $|\Delta\sigma_{\rm sky}|<1.0\times10^{-9}\ {\rm Jy\,pixel^{-1}}$, approximately corresponding to $24.4\ {\rm mag\,arcsec^{-2}}$.

\begin{figure}
    \centering
    \includegraphics[width=\columnwidth]{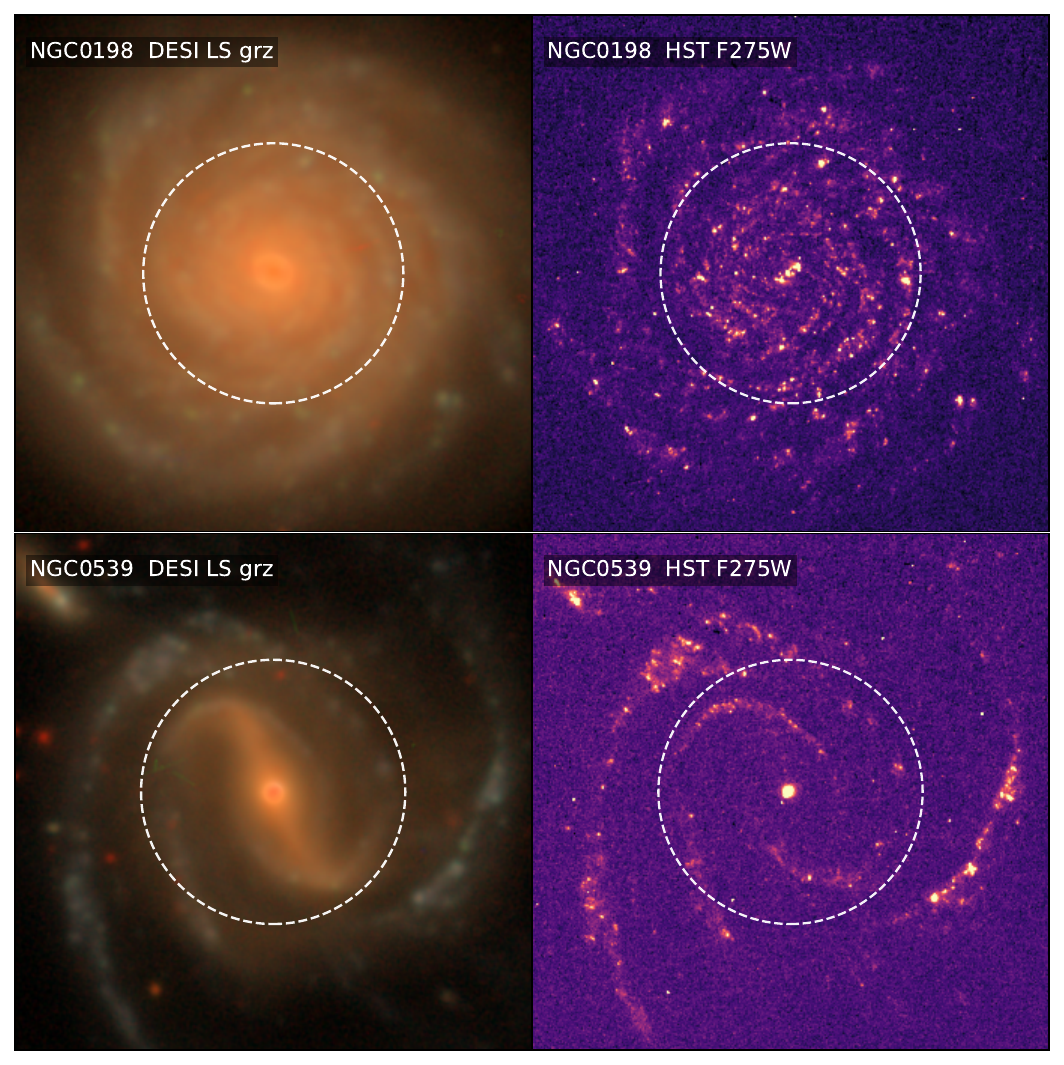}
	\caption{Representative examples of a matched control galaxy (top) and a barred galaxy (bottom). The left panels show the DESI $grz$ composite images, and the right panels show the corresponding HST/WFC3 F275W images. The dashed circle marks the optical effective radius ($R_{\rm eff}$) of each galaxy.
	\label{fig:bar_examples}}
\end{figure}

Galaxies without suitable matches were excluded from the subsequent analysis. These criteria yielded a main matched sample of 70 barred galaxies and 95 distinct unbarred control galaxies. Their stellar mass and $g-r$ color distributions are shown in Figure~\ref{fig:matched_mass_color}. We used two-sample Kolmogorov--Smirnov (KS) tests to compare the distributions of the two samples. The resulting $p$-values indicate no statistically significant differences in either stellar mass or optical color.
The median background noise is $3.36^{+0.23}_{-0.06}\times10^{-9}\ {\rm Jy\,pixel^{-1}}$ for the barred galaxies and $3.34^{+0.08}_{-0.06}\times10^{-9}\ {\rm Jy\,pixel^{-1}}$ for the distinct control galaxies, where the uncertainties indicate the 16th--84th percentile ranges. A paired Wilcoxon signed-rank comparison between each barred galaxy and the mean $\sigma_{\rm sky}$ of its matched controls yields $p=0.86$, providing no evidence of a systematic difference in image depth between the two samples. Figure~\ref{fig:bar_examples} illustrates representative examples of the matched barred and control galaxies in the optical and F275W images.

Because more than one galaxy may satisfy the matching criteria, the cross-match between the two samples is not strictly one-to-one. A barred galaxy may have more than one suitable control galaxy, and an unbarred galaxy may be matched to multiple barred systems. 
Within each matched set, the control galaxies receive equal weights that sum to unity, so each barred galaxy and its matched control set contribute equally to the comparison. The construction of the matched-control profiles is described in Section~\ref{sec:spatial_profiles}.

A more restrictive matching procedure was adopted for the clump luminosity and size distribution functions (LDFs and SDFs, respectively) presented in Section~\ref{sec:LDF_SDF}. These measurements are particularly sensitive to physical resolution because PSF blending can alter the observed clump luminosities, radii, and turnover scales. We therefore added an additional criterion of $|\Delta s_{\rm PSF}|\leq10$ pc, where $s_{\rm PSF}$ is the physical scale corresponding to the F275W PSF at the distance of each galaxy. This requirement also limits the pairwise redshift differences to $|\Delta z|\lesssim0.007$ in our sample. Repeating the analysis with the additional requirement $|\Delta z|\leq0.005$ does not change the conclusions. This stricter selection yielded 46 barred galaxies with acceptable matches and 53 distinct unbarred control galaxies.

\section{Clump identification and measurements}
\label{sec:method}

\subsection{Clump detection}
To extract the star-forming clumps, we applied the Python package \texttt{astrodendro} \citep{Goodman2009} to the HST/F275W images. This algorithm identifies hierarchical structures through dendrogram analysis, and is well suited for resolving star-forming regions with irregular morphologies and nested substructures \citep[e.g.,][]{Gao2023, Giunchi2023, Oakes2025}. Prior to clump extraction, obvious foreground stars and background sources were masked following the procedure described in Section~\ref{sec:Optical}. 

For each image, the global background level and its pixel-to-pixel dispersion $\sigma_{\rm sky}$ were estimated from the unmasked pixels using iterative $3\sigma$-clipped statistics. The background was subtracted before constructing the dendrogram. For reporting and sample comparison, $\sigma_{\rm sky}$ was converted to flux density per pixel using the image-specific \texttt{PHOTFNU} calibration.

For clump detection, \texttt{astrodendro} requires three input parameters: the minimum detection threshold (\texttt{min\_value}), the minimum contrast between adjacent structures (\texttt{min\_delta}), and the minimum number of connected pixels (\texttt{min\_npix}). In this work, we adopted \texttt{min\_value}$=3\sigma_{\rm sky}$, \texttt{min\_delta}$=1\sigma_{\rm sky}$, and \texttt{min\_npix}$=5$ pixels. 
We assessed the sensitivity to the adopted contrast threshold by repeating the clump identification and spatial-profile analysis with the more conservative value \texttt{min\_delta}$=3\sigma_{\rm sky}$, while keeping all other detection, selection, and matching criteria unchanged. The results of this test are presented in Appendix~\ref{app:min_delta}.
Because all F275W images have the same angular pixel scale of $0.04''$, \texttt{min\_npix}$=5$ imposes a uniform minimum angular footprint. Its physical scale nevertheless varies with galaxy distance. Within the matched sample, five pixels correspond to a minimum projected area of approximately $3.6\times10^{2}$--$2.1\times10^{4}\ {\rm pc}^{2}$, or an area-equivalent radius of about 11--82 pc. We assess the resulting resolution-dependent selection effect using stricter physical-resolution matching and a common physical-footprint cut in Appendix~\ref{app:resolution_test}.

\begin{figure}
    \centering
    \includegraphics[width=\columnwidth]{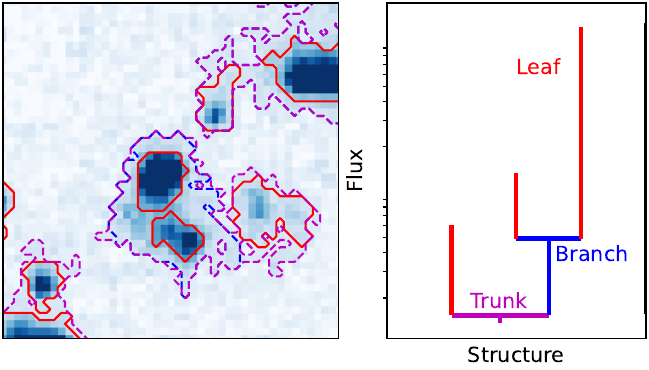}
	\caption{ Example of the dendrogram decomposition produced by \texttt{astrodendro}. The left panel shows the hierarchical segmentation of a star-forming region, where contours of different colors represent different levels of the decomposition. The right panel shows the corresponding dendrogram consisting of trunk, branch, and leaf structures. 
	\label{fig:clump_example}}
\end{figure}

Figure~\ref{fig:clump_example} presents a representative example of the dendrogram decomposition and the identified leaf nodes adopted in this work. Throughout this work, we use only the terminal structures (``leaf’’ nodes) of the dendrogram, which represent individual star-forming clumps without further resolved substructures. 

\begin{figure*}
    \centering
    \includegraphics[width=0.8\textwidth]{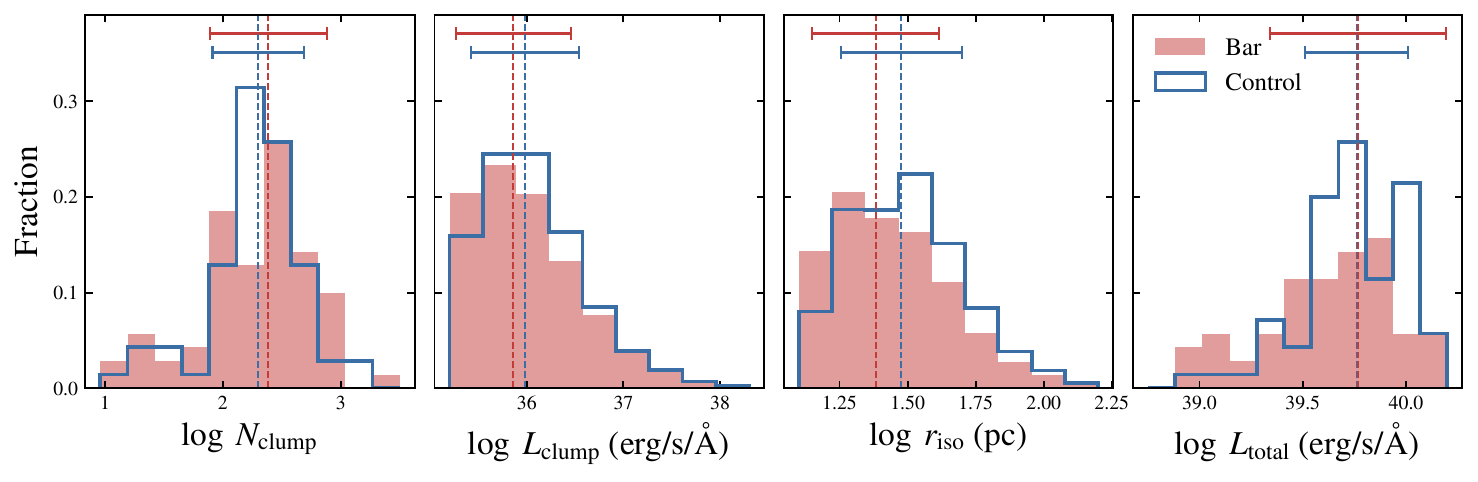}
	\caption{Fractional distributions of the basic properties of the detected star-forming clumps and their host galaxies for the barred (red filled) and matched control (blue outline) galaxies. From left to right, the panels show the number of detected clumps per galaxy ($N_{\rm clump}$), individual-clump F275W luminosity density ($L_{\rm clump}$), observed area-equivalent isophotal radius ($r_{\rm iso}$), and total galaxy F275W luminosity density ($L_{\rm total}$). The first and fourth panels show galaxy-level distributions, whereas the middle two panels show distributions for individual clumps. Vertical dashed lines mark the medians, and the horizontal error bars at the top indicate the corresponding standard deviations ($1\sigma$).
	\label{fig:clump_properties}}
\end{figure*}

\subsection{Clump photometry and size measurements}
\label{subsec:clump_photometry}
For each dendrogram leaf, the clump count rate was obtained by integrating the background-subtracted F275W signal over its segmentation footprint. The corresponding background noise was measured over an area equal to that of the clump, and the signal-to-noise ratio was defined as the integrated clump signal divided by this background uncertainty. Only clumps with ${\rm S/N}\geq3$ were retained, yielding a final catalog of 95,486 UV-bright clumps.

The clump fluxes were corrected for Galactic foreground extinction using the dust maps of \citet{Schlafly2011} and the \citet{Fitzpatrick1999} extinction curve with $R_V=3.1$. The corrected count rates were converted to F275W monochromatic luminosity densities using the \texttt{PHOTFLAM} calibration keyword in the image header and the luminosity distance of each galaxy. 

We characterized clump size using the area-equivalent isophotal radius, $r_{\rm iso}=\sqrt{A/\pi}$, where $A$ is the projected physical area enclosed by the clump segmentation footprint. This definition represents an irregular footprint by the radius of a circle with the same area and does not assume a circular clump morphology. For clumps over a comparable size range, \citet{Giunchi2023} found that the isophotal radius is typically about twice the PSF-corrected profile-based core radius, although the exact relation depends on the detection threshold, background treatment, and spatial resolution. We did not apply a direct PSF correction to $r_{\rm iso}$ because the area-defined isophotal radius does not have a unique quadrature correction. The reported sizes are therefore observed projected isophotal radii rather than intrinsic clump radii.

Figure~\ref{fig:clump_properties} summarizes the basic observed properties of the detected clumps and their host galaxies. The barred and matched control samples show broadly overlapping distributions in the number of clumps per galaxy, individual-clump F275W luminosity, and total galaxy F275W luminosity. A modest shift toward smaller median $r_{\rm iso}$ is visible in the barred sample, although the two size distributions still span similar ranges and overlap substantially. Overall, no large systematic difference is evident in the global UV output or the basic observed properties of the two clump samples. More stringent comparisons of LDFs and SDFs, based on matching the physical F275W resolution, are presented in Section~\ref{sec:LDF_SDF}.

\section{Results}
\label{sec:analysis}

\subsection{Composite maps of recent star formation}
As a first qualitative comparison, we constructed composite F275W maps for the barred and matched control samples. Before stacking, each galaxy was deprojected to a face-on orientation, normalized by its median F275W surface brightness, and rescaled by its optical effective radius ($R_{\rm eff}$). For the barred galaxies, the images were further rotated so that the bar major axes were aligned vertically.

\begin{figure*}
    \centering
    \includegraphics[width=0.8\textwidth]{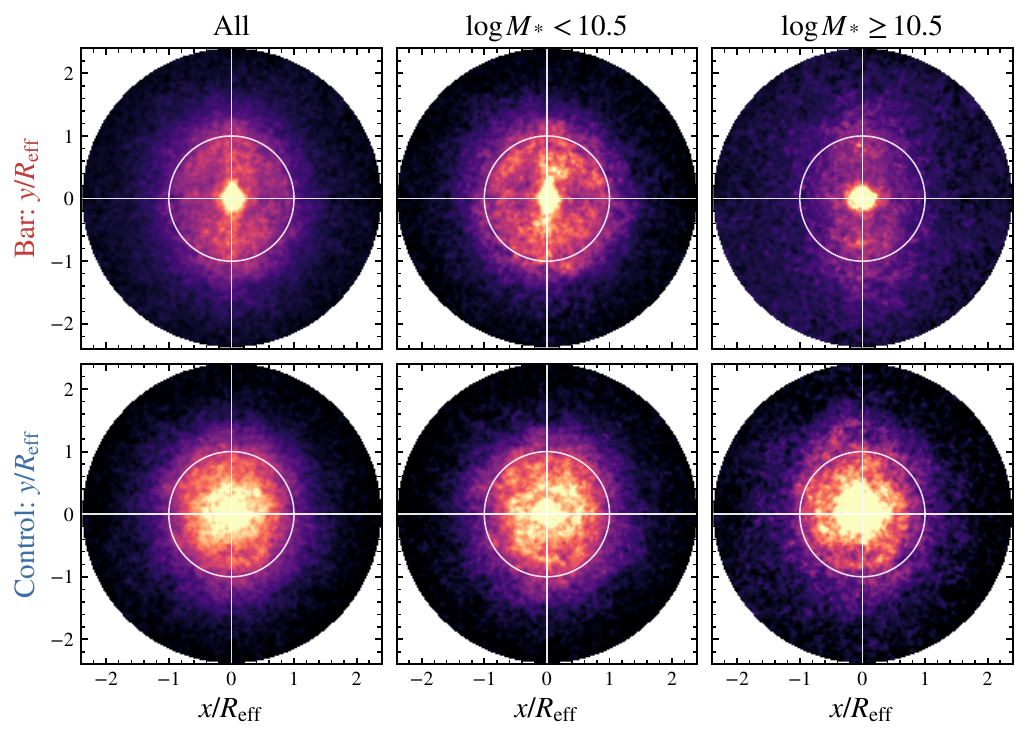}
	\caption{ Median-stacked HST/F275W images of the barred (top) and matched control (bottom) galaxies. From left to right, the panels show the full sample, the low-mass subsample ($\log(M_{\star}/M_{\odot}) < 10.5$), and the high-mass subsample ($\log(M_{\star}/M_{\odot}) \geq 10.5$). The white circle marks $R_{\rm eff}$, and the image axes are expressed in units of $R_{\rm eff}$.
	\label{fig:stacked_maps}}
\end{figure*}

Figure~\ref{fig:stacked_maps} presents the median-stacked F275W intensity maps for the full sample and for two stellar-mass bins ($\log(M_{\star}/M_{\odot})<10.5$ and $\geq10.5$). Compared with the matched control galaxies, the barred galaxies exhibit a more elongated distribution of F275W emission along the bar major axis and a clear centrally concentrated morphology. The control galaxies display a more nearly axisymmetric and less centrally concentrated distribution. These differences are visible in both stellar-mass bins, although their prominence varies with stellar mass.

The composite maps suggest that stellar bars modify the spatial distribution of recent star formation. In the following sections, we quantify these differences using the radial and azimuthal distributions of individual star-forming clumps.

\subsection{Spatial distribution of star-forming clumps}
\label{sec:spatial_profiles}
To quantify how stellar bars affect the spatial distribution of recent star formation, we constructed radial and azimuthal profiles of the UV-bright clumps in the barred and matched control galaxies. The radial profiles used 14 equal-width bins over $0\leq R/R_{\rm eff}\leq2$. The azimuthal profiles used 15 bins of width $12^\circ$ over $0^\circ$--$180^\circ$, including clumps within $R\leq2R_{\rm eff}$. The same bin edges were used for the barred and control samples.
We considered four complementary diagnostics:

\begin{enumerate}
    \item \textit{Clump number fraction.} For each galaxy, the number of clumps in a given bin, $N_j$, was normalized by the total number of retained clumps, $N_{\rm tot}$. The radial profile was further divided by the dimensionless annular area,
    $A_j=\pi\left[\left(\frac{R_{j,\rm out}}{R_{\rm eff}}\right)^2-
    \left(\frac{R_{j,\rm in}}{R_{\rm eff}}\right)^2\right]$, where $R_{j,\rm in}$ and $R_{j,\rm out}$ are the inner and outer galactocentric radii of the $j$th annulus, respectively.

    \item \textit{Total clump F275W luminosity.} The luminosities of all clumps in each bin were summed, $L_j=\sum_{i\in j}L_i$.
    For the radial profile, $L_j$ was divided by $A_j$. No additional area normalization was applied to the azimuthal profile because all azimuthal bins have the same angular width and radial coverage.

    \item \textit{F275W luminosity fraction.} For each bin, we calculated
    $f_{L,j}=\frac{\sum_{i\in j}L_i}{L_{\rm F275W,gal}}$,
    where $L_{\rm F275W,gal}$ is the integrated F275W luminosity of the host galaxy. The radial luminosity-fraction profile was also normalized by $A_j$.

    \item \textit{Median clump luminosity surface density.} For each clump, we defined $\Sigma_{L,i}=\frac{L_i}{\pi r_{{\rm iso},i}^{2}}$,
    where $r_{{\rm iso},i}$ is its area-equivalent isophotal radius. The median $\Sigma_{L,i}$ of the clumps in each bin was then used to characterize their typical UV luminosity surface density.
\end{enumerate}

For each diagnostic, we first calculated the profile of each galaxy separately. The matched-control profile for each barred galaxy was then obtained by taking the arithmetic mean of its available control profiles in each bin, in linear units. For the clump number fraction, each control profile was normalized by that galaxy's own total clump count before averaging.

\begin{figure*}
    \centering
    \includegraphics[width=1.0\textwidth]{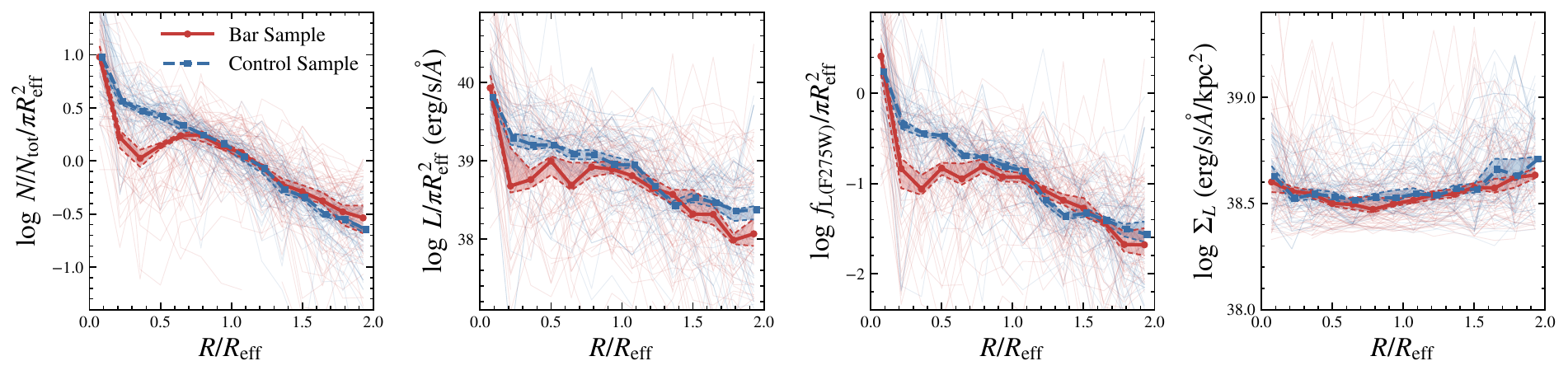}
    \includegraphics[width=1.0\textwidth]{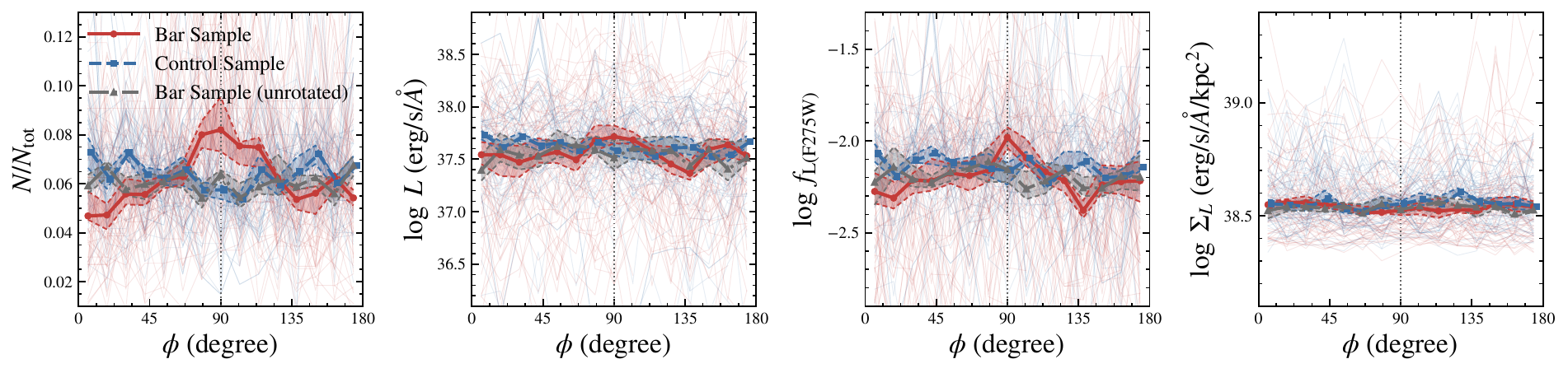}
	\caption{Radial (upper panels) and azimuthal (lower panels) distributions of star-forming clumps in the barred (red) and matched control (blue) samples. From left to right, the panels show the clump number fraction ($N/N_{\rm tot}$), total clump F275W luminosity ($L_{\rm F275W}$), F275W luminosity fraction ($f_{L({\rm F275W})}$), and median clump luminosity surface density $\Sigma_L$. The former three parameters were calculated by summing all clumps within each radial or azimuthal bin. For the radial profiles, they were further normalized by the corresponding annular area in units of $R_{\rm eff}^{2}$. The luminosity surface density of an individual clump was defined as $\Sigma_{L,i}=L_i/A_i$, where $L_i$ and $A_i$ are the F275W luminosity and projected area of an individual clump, respectively. Each thin red curve shows the profile of an individual barred galaxy, while the corresponding thin blue curve shows the bin-by-bin average profile of its matched controls. The thick red and blue curves represent the medians of the barred-galaxy profiles and their corresponding matched-control averages, respectively. Shaded regions show pointwise bootstrap uncertainty intervals, with their upper and lower boundaries marked by thin dashed lines in the corresponding colors.
   For the aligned barred-sample profiles in the lower panels, each galaxy has been rotated so that its bar major axis lies at $\phi=90^\circ$, indicated by the vertical dotted lines. The gray dash-dotted curves show the median profiles of the same barred galaxies before applying the bar-position-angle rotation, with $\phi$ retained in its original coordinate frame.
	\label{fig:composite_profiles}}
\end{figure*}

In Figure~\ref{fig:composite_profiles}, the thick curves show the median profiles of the barred galaxies and their corresponding matched-control averages. The shaded regions show bootstrap uncertainty intervals for the median profile values in each bin. We generated 400 resamples by drawing barred-galaxy indices with replacement, retaining the corresponding matched-control average profile for each selected galaxy. Individual clumps were not resampled, and both the matching and the control averages were held fixed. Each resample used the same indices across all bins. We recalculated the barred and control medians and took the 16th and 84th percentiles of their bootstrap distributions as the interval limits in linear units. For logarithmic panels, the uncertainties were converted using first-order error propagation. These intervals do not explicitly account for correlations between matched sets that share control galaxies.

The radial distributions in the upper panels of Figure~\ref{fig:composite_profiles} are shown as a function of normalized galactocentric radius, $R/R_{\rm eff}$. Compared with the unbarred sample, the barred galaxies show systematically different radial distributions. The most noticeable feature is a deficit of clumps at intermediate radii ($R\sim0.25$–$0.75R_{\rm eff}$), broadly corresponding to the bar region. In the deepest individual radial bins, the area-normalized clump number fraction, total F275W luminosity, and F275W luminosity fraction are approximately $0.5$ dex lower in barred galaxies, with the differences approaching $0.6$ dex in some luminosity-based bins. When integrated over the full intermediate radial range, the median paired barred-minus-control offsets are $-0.20$, $-0.25$, and $-0.26$ dex for these three diagnostics, respectively.

Toward the typical bar-end region ($0.75<R/R_{\rm eff}<1.25$), the barred profiles recover toward the control profiles. The corresponding region-integrated offsets range from approximately $-0.09$ to $+0.07$ dex, indicating broadly similar values between the two samples. The central differences are more dependent on the adopted diagnostic. Within $R<0.25R_{\rm eff}$, the median regional offsets are $-0.07$ dex in clump number fraction, $-0.01$ dex in total clump F275W luminosity, $+0.10$ dex in luminosity fraction, and $+0.02$ dex in median clump luminosity surface density. 

For the azimuthal distributions in the lower panels of Figure~\ref{fig:composite_profiles}, the clump position angles in each barred galaxy were shifted so that the bar major axis lies at $\phi=90^\circ$. For the unbarred controls, azimuthal angles are measured in the deprojected disk plane relative to each galaxy's own major axis, with no additional rotation. The angles are folded onto $0^\circ\leq\phi\leq180^\circ$, so the control-galaxy major and minor axes correspond to $\phi=0^\circ$ (or $180^\circ$) and $90^\circ$, respectively. The same azimuthal bins and radial selection are used for both samples. As a null comparison, we also calculated the barred-galaxy profiles without applying the bar-position-angle rotation.

The azimuthal distributions show a similarly clear signature of the stellar bars. While the control galaxies exhibit nearly axisymmetric and flat azimuthal distributions, the aligned barred galaxies display a localized maximum around $\phi=90^\circ$, corresponding to the bar major axis. In the bins nearest $\phi=90^\circ$, the clump number fraction, total F275W luminosity, and F275W luminosity fraction are approximately $0.15$--$0.20$ dex higher than in the matched controls. The same barred galaxies show no coherent peak at $\phi=90^\circ$ and are broadly similar to the matched controls when their original position angles are retained rather than aligned by the bar major axis. This null comparison indicates that the azimuthal excess is associated with the bar orientation rather than with a preferred angular direction common to the sample.

In contrast, the median clump luminosity surface density remains broadly consistent between the two samples along either the radial or azimuthal directions. The similarity between the two samples indicates that the median clump luminosity surface density changes little with the presence of a bar.

Table~\ref{tab:compact_all_medians} summarizes the regional barred--control differences and the corresponding nominal two-sided Wilcoxon signed-rank $p$ values. These values are the measurements from the median logarithmic offsets for the full sample and the mass- and bar-property subsamples, and quantify the magnitude and direction of the regional differences. For the full sample, the bar-interior deficits in all three number- and luminosity-based diagnostics have nominal $p<10^{-3}$, while the bar-aligned number-fraction excess has $p<10^{-6}$. No regional difference in median clump luminosity surface density reaches nominal significance ($p>0.20$). These tests are subject to the limitations described in Appendix~\ref{app:Regional_differences}.

The main spatial trends persist when the analysis is repeated with \texttt{min\_delta}$=3\sigma_{\rm sky}$ (Appendix~\ref{app:min_delta}). Additional tests restrict the sample to galaxies matched within 10~pc in physical PSF scale or impose a common minimum physical clump footprint within each matched set (Appendix~\ref{app:resolution_test}). These tests retain the luminosity-based deficit within the bar region and the azimuthal concentration in clump number, although the radial number-fraction deficit weakens under the common-footprint selection.

\subsection{Dependence on host galaxy and bar properties}
\subsubsection{Stellar mass dependence}
We next examine whether the spatial redistribution of star-forming clumps depends on the stellar mass of the host galaxy. The sample is divided into low-mass ($\log(M_{\star}/M_{\odot})<10.5$) and high-mass ($\log(M_{\star}/M_{\odot})\geq10.5$) subsamples. Figure~\ref{fig:composite_mass_split} compares the corresponding radial and azimuthal distributions between these two groups.

The mass-split profiles suggest that bar-related redistribution is more pronounced in massive galaxies, while the control galaxies show broadly similar distributions in both low- and high-mass bins. Relative to their matched controls, high-mass barred galaxies show a deeper and broader deficit of clump counts and luminosities within the bar region, along with a stronger azimuthal concentration along the bar major axis. 
Within the bar-interior region, the median offsets from the control sample range from $-0.30$ to $-0.50$ dex in the area-normalized clump number fraction, total F275W luminosity, and F275W luminosity fraction for the high-mass barred galaxies, compared with $-0.06$ to $-0.15$ dex at low mass.
The azimuthal difference is clearest in the clump number fraction, where the contrast between the bar-aligned sector ($\sim 90^\circ$) and off-axis sectors is approximately $0.20$ dex at high mass and $0.10$ dex at low mass. By comparison, the median clump luminosity surface density remains broadly consistent with that of the matched controls in both mass bins. 

\subsubsection{Dependence on bar properties}
\begin{figure}
    \centering
    \includegraphics[width=\columnwidth]{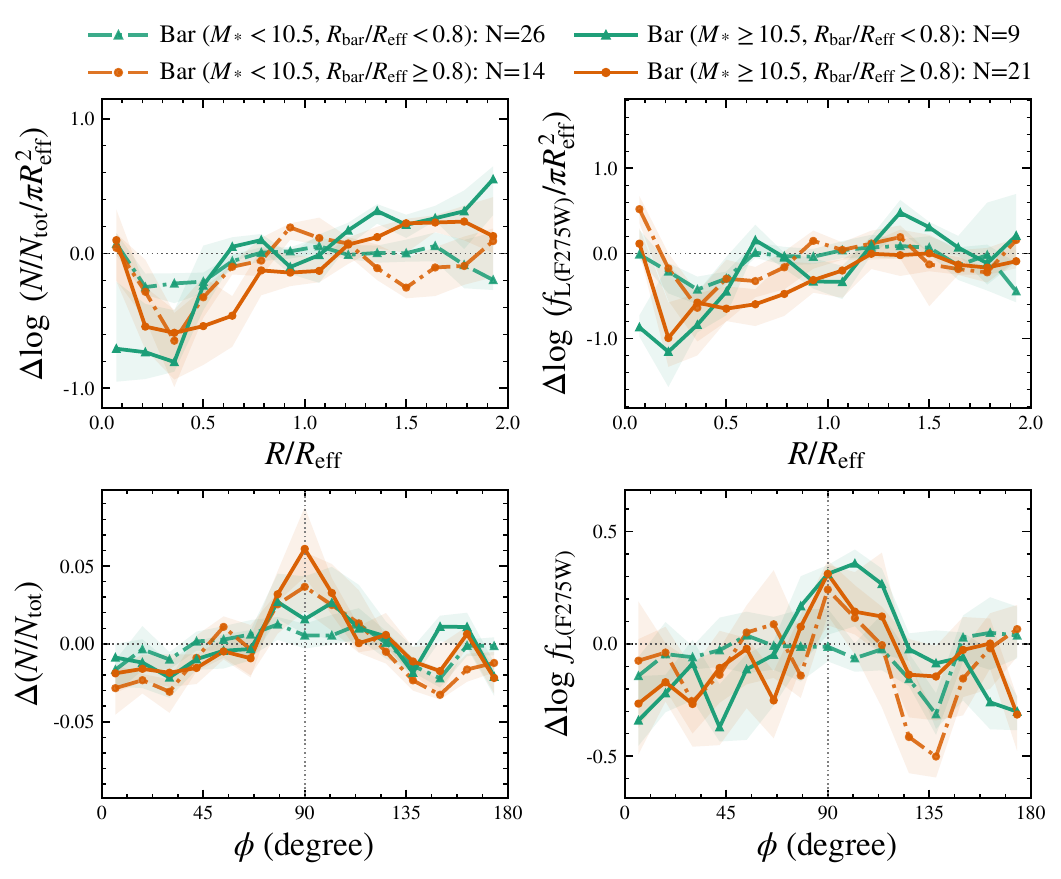}
	\caption{Bar-minus-control differences in the radial and azimuthal clump profiles for subsamples divided by stellar mass and normalized bar length. Barred galaxies are divided at $\log(M_*/M_\odot)=10.5$ and $R_{\rm bar}/R_{\rm eff}=0.8$. Each barred subsample is compared with the matched-control profile from the corresponding stellar-mass bin; the control galaxies have not been additionally divided by bar length. From upper left to lower right, the panels show $\Delta\log\ (N/N_{\rm tot}/\pi R_{\rm eff}^{2})$, $\Delta\log\ (f_{L({\rm F275W})}/\pi R_{\rm eff}^{2})$, $\Delta(N/N_{\rm tot})$, and $\Delta\log\ f_{L({\rm F275W})}$. Here, $\Delta\log\ X=\log\ X_{\rm bar}-\log\ X_{\rm control}$, while $\Delta(N/N_{\rm tot})$ was calculated in linear space. The curves represent differences between the sample-median profiles, and the shaded regions indicate the corresponding 68\% bootstrap confidence intervals. The horizontal dotted line marks no difference between barred and control galaxies. In the azimuthal panels, the bar major axis is aligned at $\phi=90^\circ$, as indicated by the vertical dotted line. Sample sizes are given in the legend.
    \label{fig:composite_mass_barsize_split}}
\end{figure}

\begin{figure}
    \centering
    \includegraphics[width=\columnwidth]{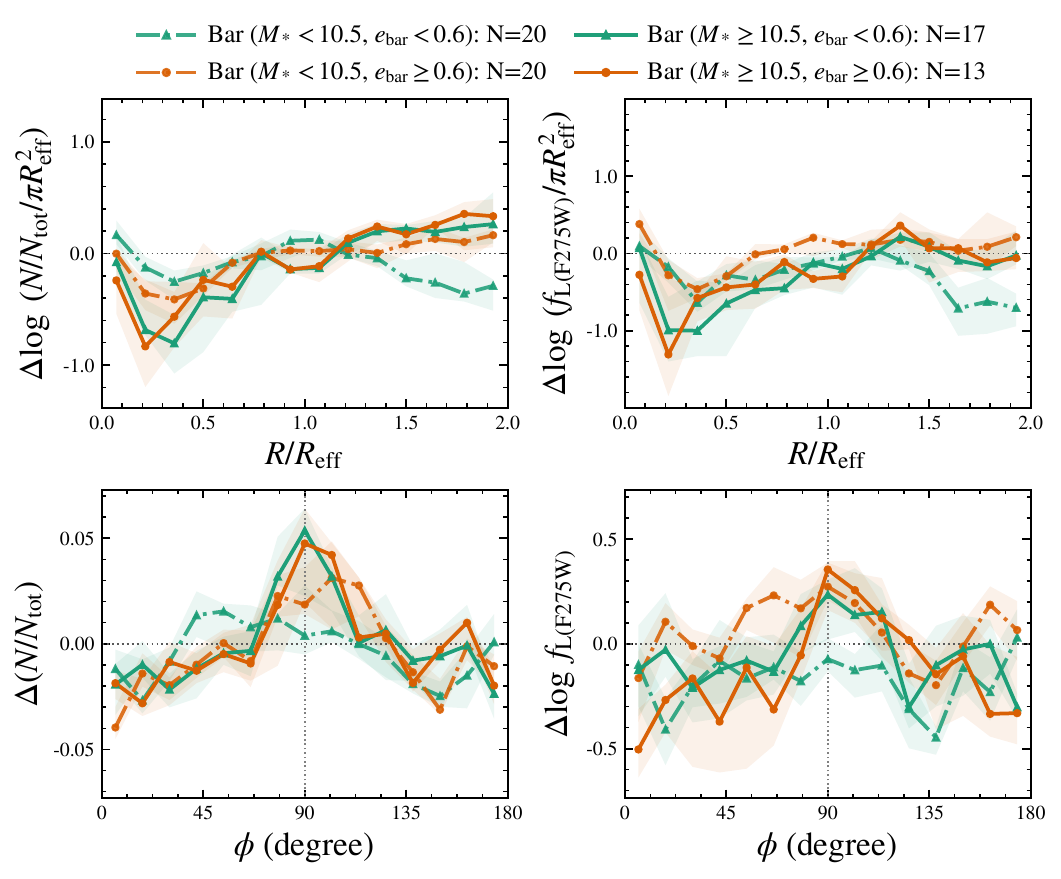}
	\caption{Same as Figure~\ref{fig:composite_mass_barsize_split} but with the barred galaxies divided by stellar mass and bar ellipticity. The adopted divisions are $\log(M_*/M_\odot)=10.5$ and $e_{\rm bar}=0.6$. Each barred subsample is compared with the matched-control profile from the corresponding stellar-mass bin. The control galaxies have not been additionally divided by bar ellipticity.
	\label{fig:composite_mass_barell_split}}
\end{figure}

We further examined whether the spatial redistribution of star-forming clumps depends on the structural properties of the stellar bars. For this purpose, the barred galaxies are divided by both stellar mass and either normalized bar length or bar ellipticity. We focused on the clump number fraction and F275W luminosity fraction, which show the clearest differences in the full profiles. Figures~\ref{fig:composite_mass_barsize_split} and \ref{fig:composite_mass_barell_split} show the difference between each barred-subsample median profile and the matched-control profile in the corresponding stellar-mass bin, with negative values indicating a deficit in the barred subsample.

Figure~\ref{fig:composite_mass_barsize_split} compares barred galaxies separated by the median normalized bar length ($R_{\rm bar}/R_{\rm eff}=0.8$). At fixed stellar mass, galaxies hosting longer bars generally show larger departures from their matched controls, although the difference varies among the diagnostics.  A clearer dependence is present at high mass. The characteristic bar-interior offsets in the longer-bar subsample are approximately $-0.50$ and $-0.60$ dex for the number and luminosity fractions, respectively, compared with approximately $-0.23$ and $-0.45$ dex for shorter bars. The corresponding radial differences between the two bar-length bins are smaller than $0.05$ dex at low mass. 
The azimuthal profiles do not show a uniformly stronger signal with increasing bar length. The contrast between the bar-aligned and off-axis sectors increases by about $0.1$ dex when moving from the short- to long-bar subsamples at low mass, but shows little change between the two bar-length bins at high mass. Overall, normalized bar length is associated more clearly with the radial than with the azimuthal differences.

\begin{figure*}
    \centering
    \includegraphics[width=\textwidth]{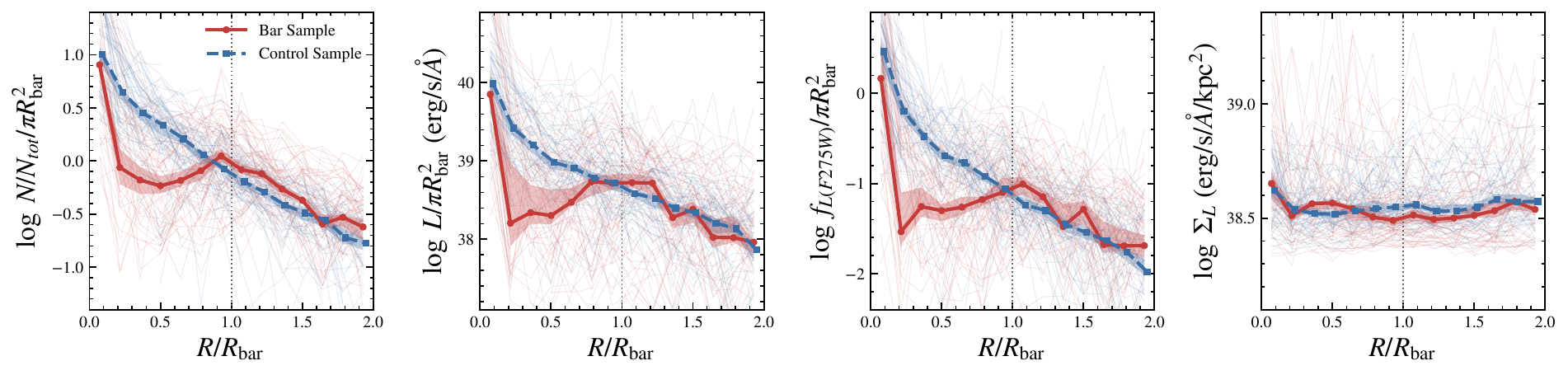}
	\caption{Similar to the top panels of Figure~\ref{fig:composite_profiles} but the radius is normalized by bar size ($R_{bar}$). For each barred galaxy, the radii of its matched control galaxies have been normalized by the same $R_{\rm bar}$. The vertical dotted line marks the nominal bar end at $R/R_{\rm bar}=1$.
	\label{fig:composite_profiles_append}}
\end{figure*}

To test whether the radial feature scales with bar length, we repeated the analysis of Figure~\ref{fig:composite_profiles} after normalizing galactocentric radius by the deprojected $R_{\rm bar}$ rather than by $R_{\rm eff}$. The matched controls of each barred galaxy were scaled by the same $R_{\rm bar}$. As shown in Figure~\ref{fig:composite_profiles_append}, the barred profiles fall farther below the control profiles ($\sim$ 1 dex compared to $0.5$ dex in Figure~\ref{fig:composite_profiles}) over parts of the inner bar region in the number- and luminosity-based diagnostics. In addition, the profiles exhibit a local recovery around $R/R_{\rm bar}\sim1$, supporting its association with the bar-end region rather than with a fixed fraction of the galaxy effective radius.

Figure~\ref{fig:composite_mass_barell_split} presents the corresponding division at $e_{\rm bar}\simeq0.6$. 
The radial dependence on ellipticity is weaker than that on normalized bar length, although the overall trends are broadly similar. At low mass, more elliptical bars show a larger number-fraction deficit but a comparable luminosity-fraction deficit. At high mass, the deficits extend over a broader radial range with increasing ellipticity.
The azimuthal behavior is also nonuniform in the two diagnostics. The azimuthal number-fraction differences generally remain below approximately $0.04$ in linear fraction. By contrast, the bar-aligned luminosity-fraction excess strengthens with increasing ellipticity, primarily in the low-mass sample. Its bar-aligned-to-off-axis contrast increases from approximately $0.07$ to $0.21$ dex at low mass but changes only slightly at high mass.

To assess the relation between host-galaxy mass and bar properties, we calculated Pearson correlations among stellar mass, normalized bar length, and bar ellipticity for the barred galaxies in the matched sample. We find that stellar mass correlates positively with normalized bar length (Pearson $r=0.35$, $p=0.0026$), while bar ellipticity shows no significant linear correlation with either stellar mass ($r=0.07$, $p=0.57$) or normalized bar length ($r=-0.01$, $p=0.94$). The correlation between mass and bar length means that their respective trends in the clump profiles may be related.

\begin{figure*}
    \centering
    \includegraphics[width=\textwidth]{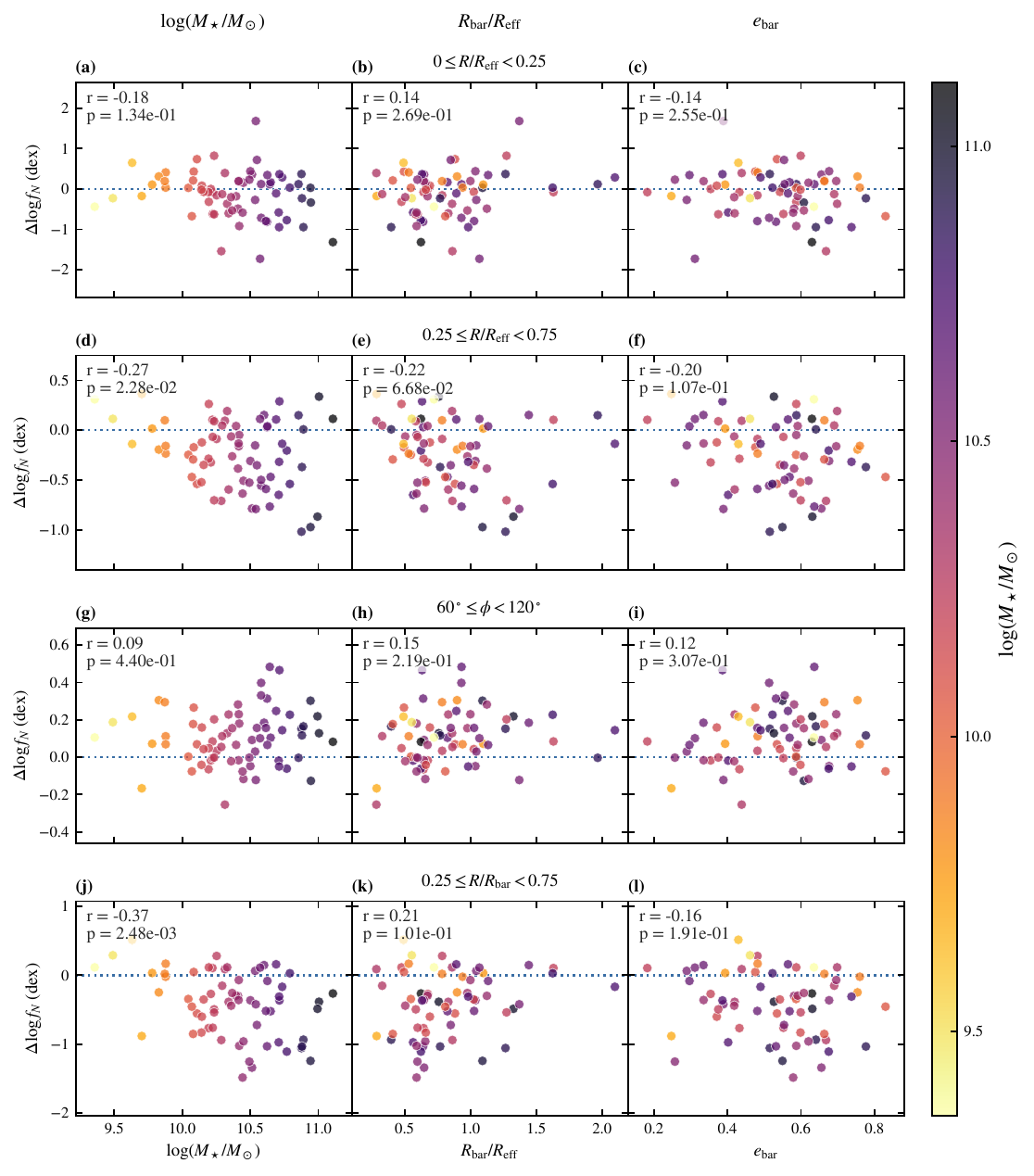}
	\caption{Regional barred--control differences in the regional clump number fraction, $f_N$, as functions of stellar mass, normalized bar length, and bar ellipticity (left to right). The regions are specified above each row. Each point represents $\Delta\log f_N=\log\ f_{N,\rm bar}-\log\ {f}_{N,\rm control}$. Points are colored by stellar mass, with a common color scale across all panels. Dotted horizontal lines mark zero difference. Each panel lists the Pearson correlation coefficient $r$ and $p$ value for the plotted points. In the bottom row, the physical bar length of each barred galaxy is used as the reference scale for that galaxy and all of its controls. Number fractions retain each galaxy's clump count within $0\leq R/R_{\rm eff}\leq2$ as the denominator.}
\label{fig:bar_control_parameter_trends}
\end{figure*}

We further examined these trends using the regional barred--control differences. Figure~\ref{fig:bar_control_parameter_trends} shows the clump number-fraction differences as functions of stellar mass, normalized bar length, and ellipticity, with the corresponding luminosity-fraction comparisons shown in Appendix Figure~\ref{fig:bar_control_parameter_trends_luminosity}. 
We used the predefined central ($R/R_{\rm eff}<0.25$), bar-interior ($0.25\leq R/R_{\rm eff}<0.75$), and bar-aligned ($60^\circ\leq\phi<120^\circ$) regions rather than individual bins. We also examined $0.25\leq R/R_{\rm bar}<0.75$, using each barred galaxy's physical bar length as the reference scale for both that galaxy and its matched controls.

Within $0.25\leq R/R_{\rm eff}<0.75$, the number-fraction differences decrease with stellar mass and normalized bar length (Pearson $r=-0.27$ and $-0.22$, respectively). The corresponding luminosity-fraction correlations are stronger ($r=-0.33$ and $-0.40$). The central and bar-aligned differences show weaker correlations with all three parameters ($|r|<0.20$). When the bar-interior region is instead defined in units of $R_{\rm bar}$, the negative correlations with stellar mass remain ($r\simeq-0.36$), whereas the correlations with normalized bar length change to $r=0.21$ and $0.04$. The apparent dependence on bar length is therefore sensitive to the radial region adopted. These comparisons do not establish independent effects of stellar mass and bar properties. Their physical implications are discussed in Section~\ref{sec:discuss}.

\subsubsection{Central recent star formation}
\begin{figure*}
    \sidecaption
    \includegraphics[width=0.7\textwidth]{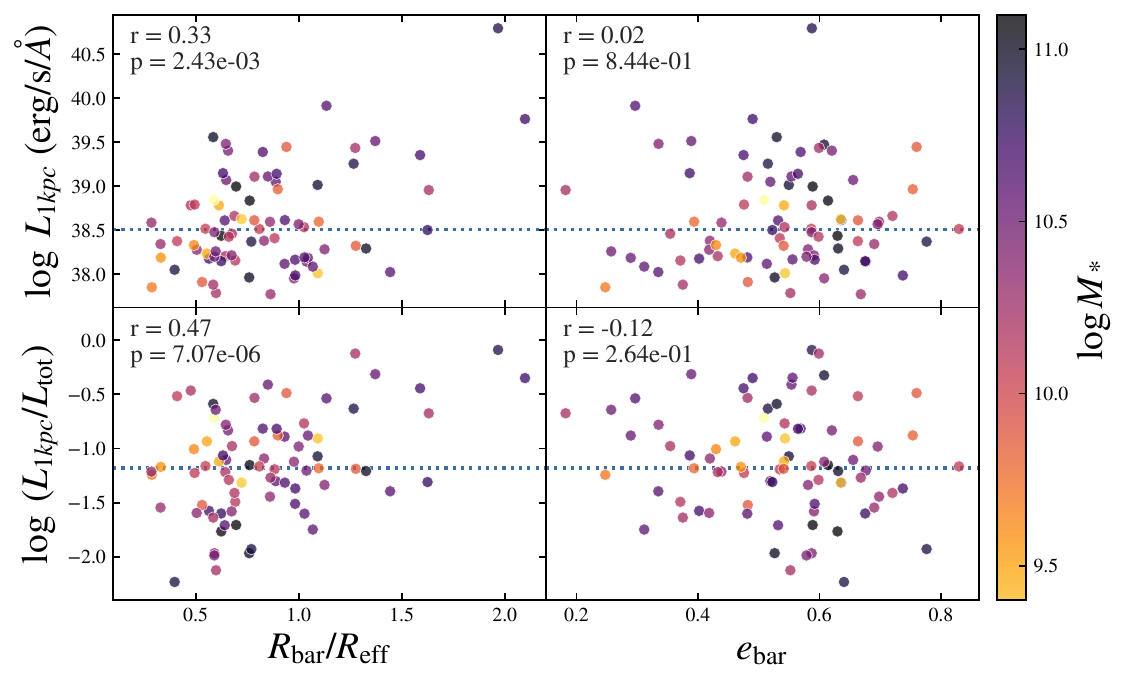}
	\caption{Relations between the central recent star formation and stellar bar properties. The upper and lower panels show the integrated F275W luminosity within the central 1 kpc, $L_{\rm 1kpc}$, and its fraction relative to the total galaxy F275W luminosity, $L_{\rm 1kpc}/L_{\rm tot}$, respectively. The left panels show the dependence on normalized bar length ($R_{\rm bar}/R_{\rm eff}$), while the right panels show the dependence on bar ellipticity ($e_{\rm bar}$). Symbols are color coded by galaxy stellar mass. The Pearson correlation coefficient ($r$) and the corresponding $p$-value are given in each panel. The horizontal dotted lines indicate the median values for the control sample.
	\label{fig:central_bar_property}}
\end{figure*}

The radial redistribution identified above suggests that stellar bars may also influence the concentration of recent star formation toward galaxy centers. To examine this possibility, Figure~\ref{fig:central_bar_property} compares the integrated F275W luminosity within the central 1 kpc, together with its fraction relative to the total F275W luminosity of each galaxy, against the normalized bar length and bar ellipticity. For each barred galaxy, we compared the central diagnostics with the median values of its matched controls. The median paired logarithmic offsets, defined as barred minus control, are $-0.10$ dex for the central F275W luminosity and $+0.02$ dex for the central luminosity fraction. Thus, the barred sample does not show a uniform enhancement of central recent star formation.

Within the barred sample, however, both central diagnostics increase with normalized bar length. The integrated central luminosity shows a moderate positive correlation with $R_{\rm bar}/R_{\rm eff}$ (Pearson $r=0.33$, $p=2.4\times10^{-3}$), while the central luminosity fraction exhibits an even stronger correlation (Pearson $r=0.47$, $p=7.1\times10^{-6}$). The relation for the luminosity fraction indicates that galaxies hosting longer bars tend to concentrate a larger fraction of their recent star formation into their central regions.

By comparison, the corresponding correlations with bar ellipticity are much weaker and are not statistically significant (Pearson $p>0.2$ for both quantities). Together with the spatial analyses presented above, these results support that normalized bar length is a more robust empirical indicator of bar-related redistribution of recent star formation than bar ellipticity.

\begin{figure*}
    \centering
    \includegraphics[width=1.0\textwidth]{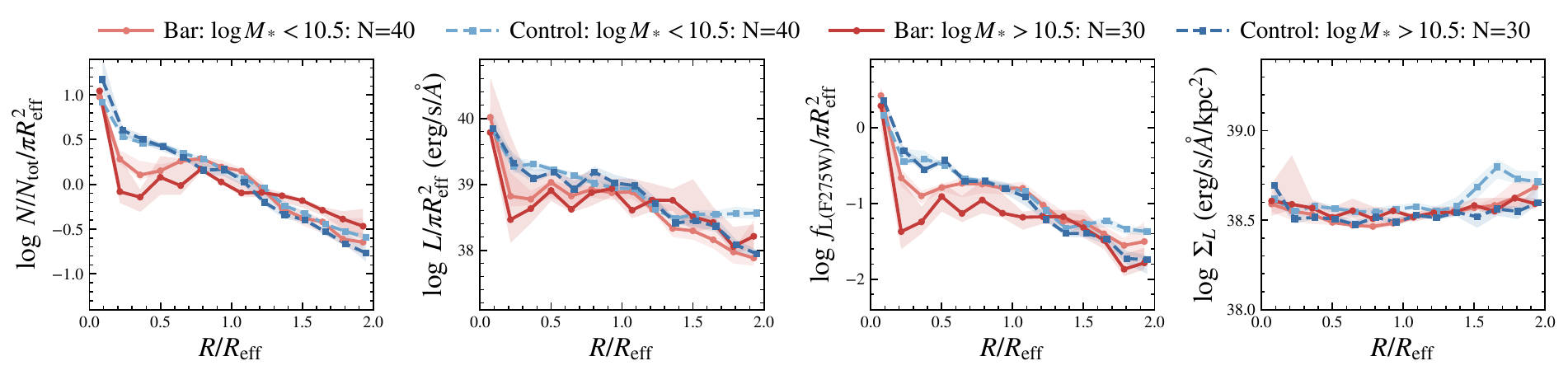}
    \includegraphics[width=1.0\textwidth]{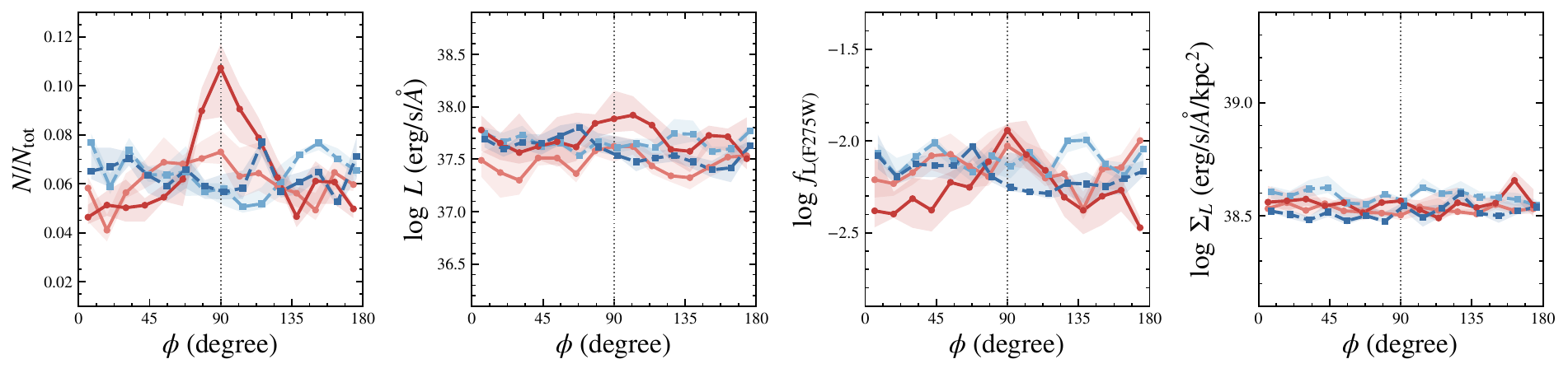}
	\caption{Same as Figure~\ref{fig:composite_profiles} but for the low-mass ($\log(M_{\star}/M_{\odot})<10.5$) and high-mass ($\log(M_{\star}/M_{\odot})\geq10.5$) subsamples.
	\label{fig:composite_mass_split}}
\end{figure*}

\begin{figure*}
    \sidecaption
    \includegraphics[width=0.7\columnwidth]{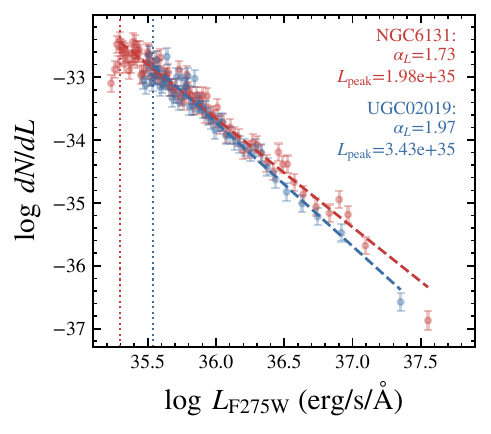}
    \includegraphics[width=0.7\columnwidth]{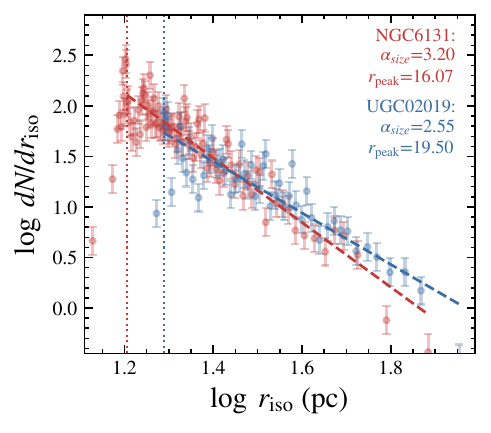}
	\caption{Representative LDFs (left panel) and SDFs (right panel) for a barred galaxy (red) and its matched control galaxy (blue). The vertical dashed lines indicate the turnover luminosity ($L_{\rm peak}$) and turnover radius ($r_{\rm peak}$). Above these turnover values, the distributions were fitted with single power-law functions, $dN/dL = K L^{-\alpha_{\rm L}}$ for the LDF and $dN/dr = K r^{-\alpha_{\rm size}}$ for the SDF, where $K$ is the normalization constant and $\alpha_{\rm L}$ and $\alpha_{\rm size}$ are the corresponding power-law slopes. The best-fitting power-law models are shown by the dashed curves.
	\label{fig:DF_fitting_examples}}
\end{figure*}

\begin{figure*}
    \centering
    \includegraphics[width=\columnwidth]{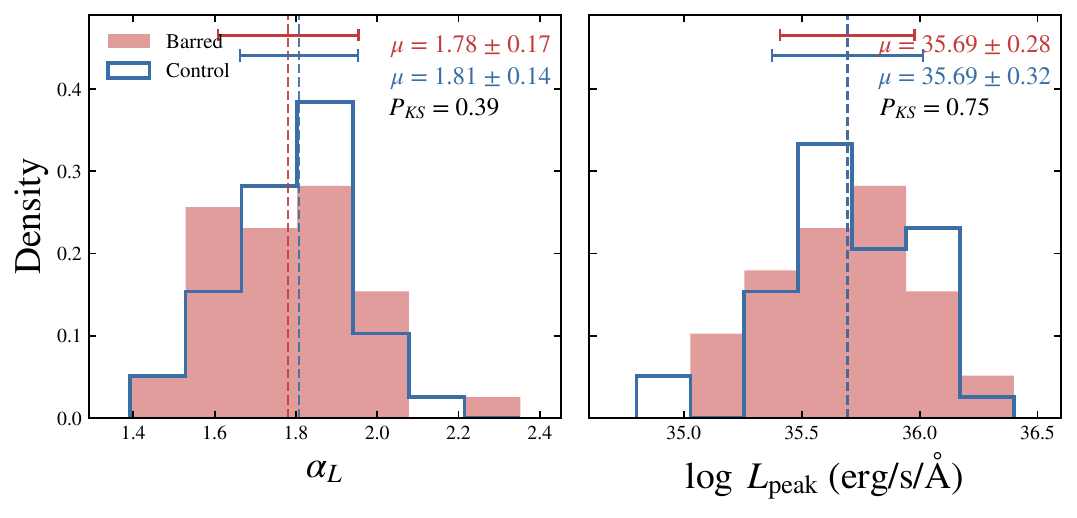}
    \includegraphics[width=\columnwidth]{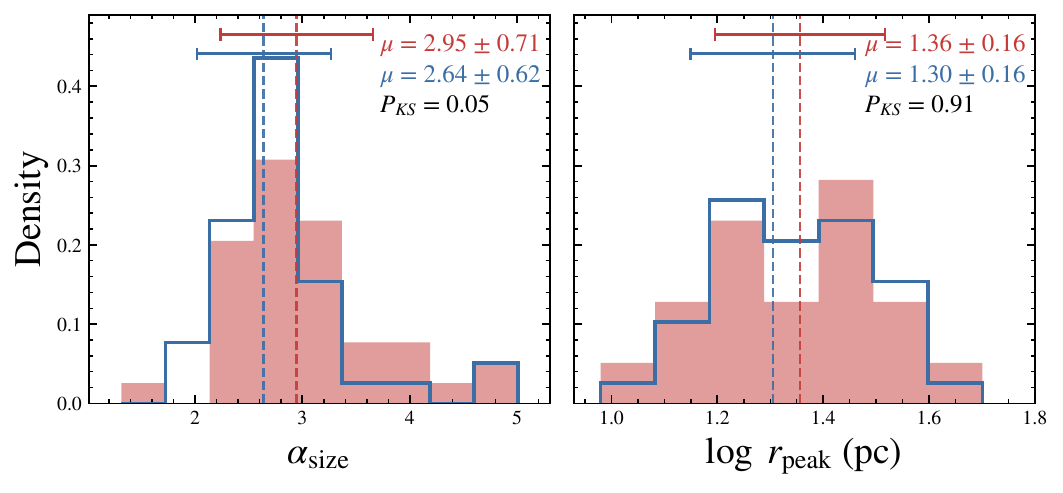}
	\caption{Comparison of the fitted LDF (two left panels) and SDF (two right panels) parameters for barred galaxies (red filled) and their matched controls (blue outline). From left to right, the panels show the LDF power-law slope ($\alpha_{\rm L}$), the turnover luminosity ($L_{\rm peak}$), the SDF power-law slope ($\alpha_{\rm size}$), and the turnover radius ($r_{\rm peak}$). The vertical dashed lines mark the median of each distribution, while the horizontal error bars at the top indicate the corresponding standard deviations. The median values ($\mu$), standard deviations, and two-sample KS-test probabilities ($P_{\rm KS}$) are reported in each panel.
	\label{fig:DF_fit_comparison}}
\end{figure*}

\subsection{Distribution functions of star-forming clumps}
\label{sec:LDF_SDF}
Finally, we examined whether stellar bars are associated with differences in the observed luminosity and size distributions of individual star-forming clumps. For this comparison, we used the more restrictive resolution-matched subsample described in Section~\ref{sec:matched_control}, comprising 46 barred galaxies and 53 distinct unbarred control galaxies. The LDFs and SDFs were constructed using adaptive, equal-occupancy bins with a target of ten clumps per bin. The variable bin widths give approximately comparable relative counting uncertainties across the distributions.

For each galaxy, we located the maximum of the binned $dN/dL$ or $dN/dr$ distribution. We tested power-law fits starting at that bin or up to two bins above it, requiring at least five bins in the fitted range. We selected the fit with the lowest reduced $\chi^2$ and defined $L_{\rm peak}$ or $r_{\rm peak}$ as the geometric mean of the endpoints of its first bin. These criteria yielded valid matched comparisons for 39 barred galaxies in each analysis, with at least one successfully fitted control per barred galaxy. When several valid controls were available, their fitted parameters were represented by the median value in the matched comparison. The fitted slopes are denoted by $\alpha_{\rm L}$ and $\alpha_{\rm size}$, respectively. Representative fits are shown in Figure~\ref{fig:DF_fitting_examples}.
 
These thresholds are not independently measured completeness limits. They depend on both the clump population and observational effects, including image depth, resolution, and source blending. Matching image depth and physical resolution reduces differential selection effects but may also contribute to the similar $L_{\rm peak}$ and $r_{\rm peak}$ values in the two samples. Their agreement alone therefore does not establish a intrinsic similarity between the clump populations, nor does the matching guarantee completeness above these thresholds.

Figure~\ref{fig:DF_fit_comparison} compares the fitted parameters for the barred and control samples. The LDF parameters are statistically consistent between the two samples. Both the luminosity slopes ($\alpha_{\rm L}$) and turnover luminosities ($L_{\rm peak}$) show overlapping distributions, with KS-test p-values of $P_{\rm KS}=0.39$ and $0.75$, respectively. These values indicate no statistically significant difference between barred and control galaxies in the luminosity distribution of their star-forming clumps.

The SDF slope distributions show a marginal difference (nominal $P_{\rm KS}\simeq0.05$), whereas the turnover-radius distributions show no statistically detectable difference ($P_{\rm KS}=0.91$). The median size power-law slope is $\alpha_{\rm size}=2.95 \pm 0.71$ for barred galaxies and $2.64 \pm 0.62$ for control galaxies. The turnover radii also overlap, with median values of $\log\ (r_{\rm peak}/{\rm pc})=1.36 \pm 0.16$ for barred galaxies and $1.30 \pm 0.16$ for control galaxies.

The barred and control samples have broadly similar observed clump luminosity and size distributions under comparable observing conditions, although intrinsic differences may remain unresolved or undetected. Together with the spatial analyses above, these results suggest that the clearest differences associated with bars lie in the spatial distribution of recent star formation rather than in the observed luminosities or sizes of individual clumps.

\section{Discussion}
\label{sec:discuss}

\subsection{Bars as regulators of the spatial distribution of recent star formation}

The main result of our work is that stellar bars reorganize the spatial distribution of recent star formation within disk galaxies. Relative to the matched control sample, the star-forming clumps in barred galaxies show a modest central enhancement, an intermediate-radius deficit, and a local recovery near the typical bar-end radius (Figures~\ref{fig:stacked_maps} and \ref{fig:composite_profiles}). Their clumps also preferentially align with the bar major axis (Figure~\ref{fig:composite_profiles}). These observational signatures consistently indicate that bars primarily reshape where recent star formation occurs.

This interpretation is consistent with the current framework of bar-driven secular evolution. Non-axisymmetric gravitational torques of bars can redistribute angular momentum and drive gas inward along the bar \citep[e.g.,][]{Regan2004, Athanassoula2013}. The resulting gas transport naturally produces central gas concentrations while reducing the gas reservoir available for star formation across much of the bar region \citep{Lopez-Coba2022, Hogarth2024, Scaloni2024}. Gas can also accumulate near bar ends, inner rings, or bar--spiral interfaces, where orbit crowding and enhanced gas density may promote gravitational collapse \citep[e.g.,][]{Diaz-Garcia2020, Neumann2020, Renu2025}. 
The positive correlation between the central F275W luminosity fraction and normalized bar length (Figure~\ref{fig:central_bar_property}) is consistent with a stronger central concentration of recent star formation in galaxies with longer bars.

 The suppression of star formation across much of the bar region, commonly referred to as the ``star formation desert'' \citep{James2009, James2015}, likely results from a combination of reduced gas supply and the hostile dynamical environment within the bar. Numerical simulations and observations suggest that strong shocks, large velocity gradients, and elevated shear induced by a strong bar can increase cloud--cloud collision velocities and inhibit the collapse of molecular clouds even when a substantial gas reservoir remains \citep{Fujimoto2020, Maeda2023, Kim2024a, Costantin2026}. 
 
At the same time, the azimuthal excess shows that UV-bright clumps preferentially lie along the bar major axis (Figure~\ref{fig:composite_profiles}). Gas compression near the leading edges of bars is one possible mechanism that could contribute to this pattern \citep[e.g.,][]{Athanassoula1992, Regan2004}, although the observed alignment alone does not establish this connection. Together with the radial differences, the azimuthal concentration is consistent with bar-driven gas redistribution influencing where recent star formation occurs. This is consistent with recent high-resolution simulations and spatially resolved observations, which likewise show that stellar bars suppress star formation over much of the bar while concentrating recent star formation along specific dynamical structures \citep[e.g.,][]{Neumann2019, Fraser-McKelvie2020b, Emsellem2026}.

The similarity of the clump properties in the two samples places an important constraint on this interpretation. Although barred and control galaxies differ in the locations of their UV-bright clumps, they have broadly consistent median clump luminosity surface densities, clump luminosity and size distribution functions (Figures~\ref{fig:composite_profiles} and \ref{fig:DF_fit_comparison}). Bars therefore appear to influence the large-scale placement of recent star formation more strongly than the observed luminosities and sizes of individual clumps after star formation has commenced. This result does not establish that the internal physics of molecular clouds is unaffected by bars, because the UV data do not directly resolve cloud-scale gas dynamics or the earliest embedded phases of star formation. It instead shows that any such differences do not produce a measurable change in the clump-scale UV properties examined here.

This picture may also help reconcile the diverse conclusions regarding the bar--star formation connection. Previous studies have reported enhanced, suppressed, or comparable star formation activity in barred and unbarred galaxies \citep[e.g.,][]{Wang2020, Geron2023}. If stellar bars mainly redistribute recent star formation without substantially changing its global level, measurements based on integrated star formation indicators may naturally yield weak or even contradictory trends. The outcome may also depend on host-galaxy mass, bar properties, the adopted star formation tracer, and the evolutionary stage sampled \citep[e.g.,][]{Diaz-Garcia2021, George2021, Mengistu2026}. Resolving individual star-forming clumps therefore provides a more direct way to isolate the dynamical influence of stellar bars on recent star formation. 

\subsection{Dependence on galaxy mass and bar properties}

The spatial redistribution associated with stellar bars is more pronounced in the high-mass galaxies of our sample (Figure~\ref{fig:composite_mass_split}). Relative to their matched controls, massive barred galaxies show a broader deficit of star-forming clumps within the bar region, together with stronger central and azimuthal concentrations. This trend does not appear to result solely from massive galaxies hosting systematically older bars, as direct bar-age estimates show no significant relation between bar age and host-galaxy stellar mass \citep{deSa-Freitas2025}. Stellar mass may instead influence how the interstellar medium (ISM) responds to the non-axisymmetric bar potential. Recent PHANGS-JWST observations identify a transition in interstellar-medium morphology near $M_{\star}\sim10^{10}\,M_{\odot}$ \citep{Emsellem2026}. Below this mass,  the ISM is uniformly clumpy and irregular regardless of bar presence, whereas more massive barred galaxies typically feature central disks or rings and a cleared bar region. This behavior resembles the stronger central excess and intermediate-radius deficit found in our high-mass subsample. In addition, lower-mass galaxies also tend to be more gas rich \citep[e.g.,][]{Saintonge2022}, and their less-ordered star-forming structures may weaken or obscure a coherent bar-related pattern. The observed mass dependence may therefore reflect changes in both the dynamical response and the spatial organization of the multiphase interstellar medium rather than a simple variation in bar torque strength.

Within the barred sample, normalized bar length shows a more systematic relation with recent star formation than bar ellipticity (Figures~\ref{fig:composite_mass_barsize_split} and \ref{fig:composite_mass_barell_split}). Within each stellar-mass bin, galaxies with longer bars exhibit stronger radial deficits, whereas the azimuthal dependence is less uniform. Across the full barred sample, normalized bar length is also positively related to the central F275W luminosity and, more clearly, to the fraction of F275W luminosity contained within the central kiloparsec (Figure~\ref{fig:central_bar_property}).

The alternative radial normalization provides an additional clue to the spatial role of the bar. When galactocentric radius is scaled by the deprojected bar length, the barred profiles remain depressed across much of the region within $R_{\rm bar}$ and show a local recovery near $R/R_{\rm bar}\sim1$ (Figure~\ref{fig:composite_profiles_append}). This feature is clearest in the clump number distribution. Its alignment with the nominal bar end is consistent with recent star formation accumulating near the bar-end region, rather than at a fixed fraction of $R_{\rm eff}$. 
This recovery may also be related to dynamical scales associated with the bar. For fast bars, the corotation radius typically lies at $R_{\rm CR}/R_{\rm bar}\simeq1$--$1.4$ \citep{Aguerri2015}. Inner rings near the bar ends have also been linked to orbital structures between the inner 4:1 resonance and corotation \citep{Patsis2003}.
The result therefore suggests that $R_{\rm bar}$ provides a useful physical scale for describing the spatial redistribution associated with stellar bars.

Unlike the bar length,  the dependence on bar ellipticity is less uniform, although massive galaxies with highly elongated bars still display a pronounced redistribution signal. The different behavior of bar length and ellipticity may reflect the distinct physical information carried by these parameters. Numerical models predict that angular-momentum exchange with the outer disk and dark matter halo can produce progressively longer and more extended bars \citep{Athanassoula2013, Fragkoudi2025}. Direct bar-age estimates likewise suggest that older bars tend, on average, to be longer and stronger \citep{deSa-Freitas2025}. Normalized bar length may therefore preserve information about the cumulative dynamical evolution of the bar, although it also depends on the structural evolution and measurement of $R_{\rm eff}$ \citep{Diaz-Garcia2016, Geron2025}.

Bar ellipticity describes bar shape rather than directly measuring its dynamical strength. Its measured value can be affected by contamination from bulge and disk light, the three-dimensional thickness of the bar, and uncertainties in deprojection \citep{Zou2014, Sellwood2020}. Ellipticity also provides only an indirect measure of the non-axisymmetric force, which depends on the full stellar mass distribution \citep{Fragkoudi2015, Kim2024b}. These effects can introduce additional scatter into its relation with recent star formation and weaken its value as a tracer of cumulative bar evolution or bar-driven torques. In the present sample, normalized bar length is therefore the more informative empirical predictor of UV-clump redistribution. 

Bar length and ellipticity describe bar geometry but do not fully characterize its dynamical strength. Complementary measures include the normalized $m=2$ Fourier amplitude, $A_2$, and the bar torque parameter, $Q_{\rm b}$, defined as the maximum tangential-to-mean-radial force ratio within the bar region \citep{Diaz-Garcia2016}. Combining these measures with resolved gas kinematics and multivariate analyses of larger samples could help distinguish the roles of stellar mass, bar geometry, and dynamical strength in shaping the observed clump distributions.

\subsection{Clump luminosity and size distributions}
\label{subsec:disc_clump_properties}

Despite their different spatial distributions, the F275W-bright clumps in barred and control galaxies have broadly similar luminosity and size distribution functions. The fitted slopes and turnover values overlap substantially between the two samples. Stellar bars therefore appear to influence where star-forming clumps occur more strongly than the observed luminosities and sizes of individual clumps.

Resolved studies show that environmental differences can be more apparent in some cloud properties than in others. PHANGS-ALMA observations reveal elevated molecular-cloud line widths and virial parameters in bars and galaxy centers compared with other disk environments \citep{Rosolowsky2021}. Similarities in other measured properties can also coexist with changes in the cloud population. \citet{Maeda2020} find that the properties of giant molecular clouds in NGC~1300 are broadly comparable across the bar, spiral arms, and bar ends, yet the flatter mass spectrum at the bar ends indicates a greater relative contribution from massive clouds. \citet{Sanchez-Menguiano2026} find that H\,\textsc{ii} regions in bars are generally more luminous in H$\alpha$ and more dust-attenuated than those in the surrounding disks. These results suggest that similar distributions in selected observables need not imply identical gas conditions or star-forming environments.

These findings are not necessarily inconsistent with our UV-clump results. Bar-driven torques, shocks, and streaming motions act across kiloparsec scales, regulating where gas is accumulated, depleted, or exposed to strong shear. The subsequent evolution of star-forming regions also depends on local cloud conditions and stellar feedback \citep[e.g.,][]{Chevance2020}. Environmental differences in molecular gas therefore need not produce equivalent differences in the resulting UV emission. CO, H$\alpha$, and F275W also trace different stages of star formation, with F275W sampling stellar populations over a broader age range than H$\alpha$. Moreover, our LDFs and SDFs combine clumps from different locations within each galaxy. Differences confined to the bar or bar ends may consequently be diluted in these galaxy-wide distributions.

The interpretation is also limited by what an F275W-selected clump represents. Its luminosity depends not only on stellar mass, but also on age, dust attenuation, and the sampling of short-lived massive stars. Its measured size depends on spatial resolution, blending, and the adopted dendrogram segmentation. A UV clump may contain several stellar clusters or associations and need not correspond one-to-one with a single molecular cloud. Resolution limitations also affect clump-size measurements in other HST studies \citep{Giunchi2023}. Although our power-law fits exclude the distributions below $L_{\rm peak}$ and $r_{\rm peak}$, these empirical thresholds do not establish completeness or remove the effects of resolution and blending.

\subsection{Limitations and future prospects}
\label{subsec:limitations}

Several limitations should be considered when interpreting our results. F275W emission traces relatively unobscured young stellar populations over timescales of up to $\sim100$ Myr, but its luminosity depends on stellar age, dust attenuation, metallicity, and recent star-formation history \citep[e.g.,][]{Kennicutt2012, Salim2020}. Embedded star-forming regions may therefore be underrepresented, particularly in galaxy centers and along bar dust lanes. Spatially varying attenuation could modify the measured radial and azimuthal profiles, although dust alone is unlikely to explain the coherent differences between the barred galaxies and their matched controls. Combining ultraviolet imaging with extinction-corrected recombination-line and infrared tracers would provide a more complete census of recent star formation in these dynamically complex regions.

The measured clump properties are also influenced by spatial resolution, surface-brightness sensitivity, and source segmentation. Although the archival HST images generally provide high-quality and broadly comparable imaging, potential variations in exposure depth, background noise and target distance may affect the completeness limits of the luminosity and size distributions. The closely matched $\sigma_{\rm sky}$ distributions reduce systematic differences in detection sensitivity between the barred and control samples. Nevertheless, the remaining galaxy-to-galaxy variation means that the data do not share a uniform clump-completeness limit and may affect the detection of the faintest clumps in individual galaxies.

Source blending introduces an additional uncertainty, particularly in crowded galaxy centers, where several star-forming regions may be identified as a single UV clump. Preferential blending in these regions can reduce the detected central clump count and lower its fraction of the galaxy-wide total, weakening an intrinsic central enhancement. The effect on the barred--control comparison depends on the relative degree of crowding and the radial distribution of blending in the two samples. Comparable spatial resolution does not necessarily imply comparable crowding. The central clump-number profiles should therefore be interpreted as distributions of detected UV structures, rather than a complete census of individual star-forming regions. The integrated F275W luminosity within the central 1 kpc provides a complementary diagnostic because it does not depend on how the emission is divided into clumps, although it does not quantify the number of unresolved components. The measured clump sizes and luminosities therefore represent properties at the effective resolution of the observations rather than those of individual stellar clusters.

These resolution and completeness limits may obscure subtle environmental differences even under the uniform detection and signal-to-noise criteria adopted here. In addition, bar length and ellipticity are structural proxies rather than direct measurements of bar torque or pattern speed. This limits their ability to distinguish between different stages of bar evolution. The archival sample selection additionally limits the generality of our conclusions. The relatively small number of galaxies in the subsamples split simultaneously by stellar mass and bar properties also increases the uncertainty of the trends found in these bins.

Future observations should connect the UV clumps directly to the gas and to younger, dust-obscured phases of star formation. Spatially resolved H\,{\sc i} and CO observations, combined with integral-field spectroscopy, can determine whether the UV deficit within bars reflects gas depletion or reduced star-formation efficiency, while also constraining gas kinematics, extinction, and metallicity. 
At the same time, future wide-field ultraviolet surveys will extend resolved analyses to larger and more homogeneous galaxy samples. In particular, the China Space Station Telescope (CSST) will provide NUV and optical imaging over a planned wide survey of approximately $17\,500~{\rm deg}^{2}$ \citep{Zhan2011, CSSTCollaboration2026}. Its survey camera combines a field of view of about $1.1~{\rm deg}^{2}$ with a pixel scale of approximately $0\farcs074$ and an imaging requirement of an 80\% encircled-energy radius no larger than about $0\farcs15$. Therefore, CSST will enable systematic measurements of recent star formation across substantially larger samples spanning broad ranges of stellar mass, environment, and bar properties. Joint analyses combining the survey scale of CSST with high-resolution ultraviolet imaging, resolved gas measurements, and integral-field spectroscopy will provide a direct route to determining how stellar bars regulate the location and efficiency of star formation within disk galaxies.

\section{Summary}
\label{sec:summary}

We investigated how stellar bars influence the spatial distribution and observed properties of recent star formation using archival HST/WFC3 F275W imaging. Our final sample contained 70 barred galaxies and 95 matched unbarred control galaxies selected to have similar stellar masses, optical colors, redshifts, and F275W background-noise levels. Individual UV-bright star-forming clumps were identified uniformly with a dendrogram-based method, yielding a catalog of 95\,486 UV-bright structures. We compared their radial and azimuthal distributions, central F275W emission, and luminosity and size distribution functions. Our main results are summarized as follows.

\begin{enumerate}
    \item Stellar bars produce a clear spatial redistribution of recent star formation. 
    Compared with the matched controls, barred galaxies show a deficit of recent star formation over $0.25<R/R_{\rm eff}<0.75$. The region-integrated barred-minus-control offsets are approximately $-0.20$, $-0.25$, and $-0.26$ dex in the clump number fraction, total F275W luminosity, and F275W luminosity fraction, respectively. The barred profiles recover toward the controls near the typical bar-end region.
    
    \item Ultraviolet-bright clumps are preferentially concentrated along the bar major axis. Near $\phi=90^\circ$, the number- and luminosity-based diagnostics exceed those of the matched controls by approximately $0.15$--$0.20$ dex. No corresponding peak appears when the barred galaxies retain their original position angles.

    \item The bar-related redistribution of UV clumps is more pronounced in massive galaxies. High-mass barred systems exhibit a broader radial deficit and stronger central and azimuthal concentrations than their matched controls, whereas the corresponding differences are weaker in lower-mass galaxies.

    \item The normalized bar length shows a clearer association with the radial deficit and central F275W concentration than the bar ellipticity, whereas its relation with the azimuthal signal is less uniform. Normalizing the radius by $R_{\rm bar}$ places the local recovery near $R/R_{\rm bar}\sim1$, identifying bar length as a useful scale for describing the redistribution.

    \item The LDF and SDF analyses used a stricter subsample of 46 barred and 53 control galaxies. After the fitting requirements were applied, 39 valid matched comparisons remained for both the LDF and SDF. Within the sensitivity and physical resolution of the data, their fitted slopes and turnover scales showed no robust systematic differences between barred and control galaxies.
\end{enumerate}

Together, these results support a picture in which stellar bars primarily reorganize where recent star formation occurs. The observed radial and azimuthal patterns are consistent with bar-driven gas redistribution on galactic scales, while the observable properties of the resulting UV-bright clumps remain largely governed by local physical processes. Future homogeneous UV surveys, particularly those conducted with CSST, can be combined with resolved gas observations and integral-field spectroscopy to test this picture in larger samples.

\section*{Data availability}
The reduced HST/WFC3 F275W images used in this study are publicly available from MAST. A table listing their corresponding HST data identifiers are available at the CDS via \url{http://cdsweb.u-strasbg.fr/cgi-bin/qcat?J/A+A/}

\begin{acknowledgements}
We thank the referee and editor for their critical comments and instructive suggestions that significantly improved the quality of the paper.
This work is supported by the National Key Research and Development Program of China (Nos. 2025YFF0510604, 2025YFF0511001, 2022YFA1602902), and the China Manned Space Project (Nos. CMS-CSST-2025-A08, CMS-CSST-2025-A19). This project is also supported by the Young Talent Fund of Institutional Center for Shared Technologies and Facilities, National Astronomical Observatories, CAS. 

This research made use of \texttt{astrodendro}, a Python package for computing dendrograms of astronomical data (\url{http://www.dendrograms.org/}). 

This work is based on observations made with the NASA/ESA Hubble Space Telescope and obtained from the Space Telescope Science Institute. These observations are associated with programs 11359, 13025, 13782, 13943, 14779, 14785, 15105, 15166, 15186, 15647, 15980, 16223, 16239, 16287, 16657, 16741, 16874, 17179.

The DESI Legacy Imaging Surveys consist of three individual and complementary projects: the Dark Energy Camera Legacy Survey (DECaLS), the Beijing-Arizona Sky Survey (BASS), and the Mayall z-band Legacy Survey (MzLS). DECaLS, BASS and MzLS together include data obtained, respectively, at the Blanco telescope, Cerro Tololo Inter-American Observatory, NSF’s NOIRLab; the Bok telescope, Steward Observatory, University of Arizona; and the Mayall telescope, Kitt Peak National Observatory, NOIRLab. The complete acknowledgments can be found at 
\url{https://www.legacysurvey.org/acknowledgment/}.
\end{acknowledgements}

\bibliographystyle{aa}
\bibliography{main.bib}

\begin{appendix}
\section{Robustness tests}
\subsection{Sensitivity to the dendrogram contrast threshold}
\label{app:min_delta}

\begin{figure*}
    \centering
    \includegraphics[width=\textwidth]{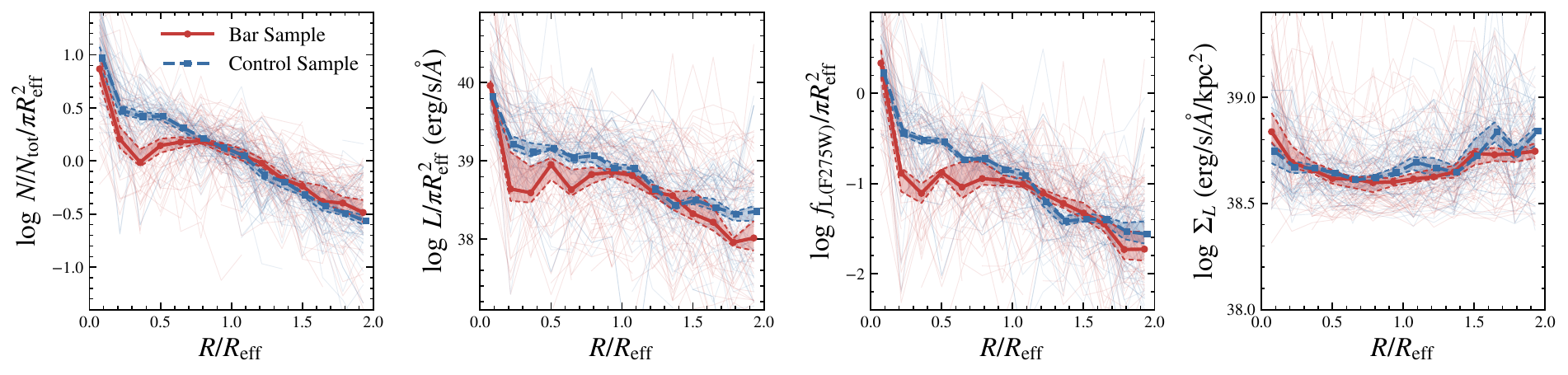}
    \includegraphics[width=\textwidth]{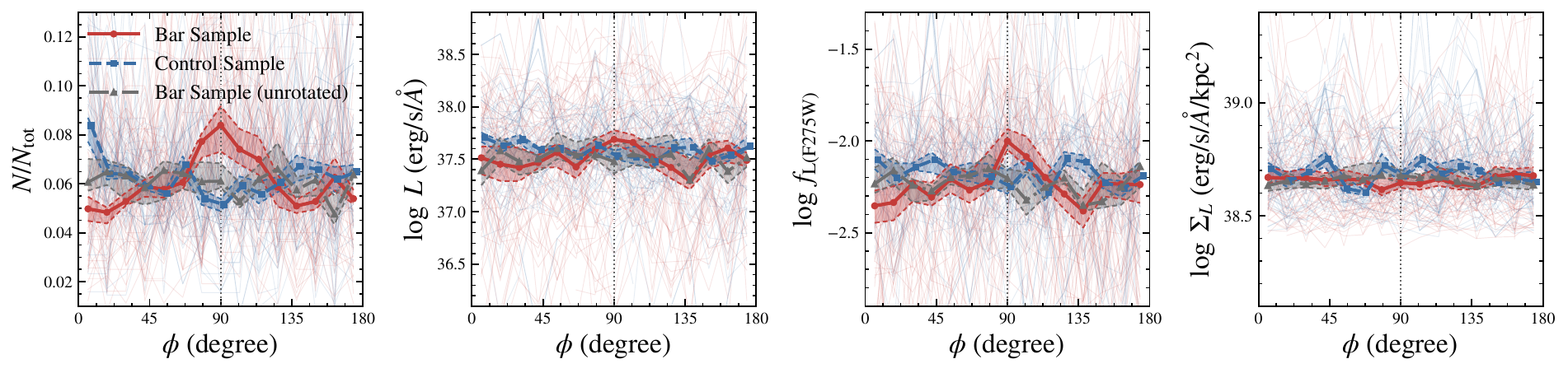}
    \caption{Same as Figure~\ref{fig:composite_profiles} but using the more conservative dendrogram contrast threshold \texttt{min\_delta}$=3\sigma_{\rm sky}$. All other detection, S/N-selection, and sample-matching criteria are unchanged.}
    \label{fig:min_delta_test}
\end{figure*}

\begin{table*}[t]
\caption{Median regional barred--control differences for the overall sample under four selections (dex).}
\label{tab:compact_robustness_medians}
\centering
\small
\setlength{\tabcolsep}{5pt}
\begin{tabular}{llrrrrrr}
\hline\hline
Selection & Diagnostic & \multicolumn{3}{c}{Radial range ($R/R_{\rm eff}$)} & \multicolumn{3}{c}{Azimuthal region ($\phi$, deg)} \\
\cline{3-5}\cline{6-8}
 & & 0--0.25 & 0.25--0.75 & 0.75--1.25 & 0--60 & 60--120 & 120--180 \\
\hline
Fiducial & $f_N$ & $-0.07$ & $-0.20$ & $0.06$ & $-0.06$ & $0.10$ & $-0.04$ \\
 & $L_{\rm F275W}$ & $-0.01$ & $-0.25$ & $-0.10$ & $-0.08$ & $0.00$ & $-0.08$ \\
 & $f_L$ & $0.10$ & $-0.26$ & $-0.01$ & $-0.07$ & $0.08$ & $-0.09$ \\
 & $\widetilde{\Sigma}_{\rm cl}$ & $0.02$ & $-0.03$ & $-0.03$ & $-0.01$ & $-0.03$ & $0.00$ \\
\hline
$\texttt{min\_delta}=3\sigma_{\rm sky}$ & $f_N$ & $0.00$ & $-0.18$ & $0.04$ & $-0.07$ & $0.09$ & $-0.04$ \\
 & $L_{\rm F275W}$ & $-0.02$ & $-0.24$ & $-0.08$ & $-0.07$ & $0.02$ & $-0.09$ \\
 & $f_L$ & $0.08$ & $-0.26$ & $-0.01$ & $-0.06$ & $0.08$ & $-0.12$ \\
 & $\widetilde{\Sigma}_{\rm cl}$ & $0.04$ & $-0.02$ & $-0.03$ & $0.01$ & $-0.02$ & $-0.01$ \\
\hline
Physical PSF matching & $f_N$ & $0.07$ & $-0.17$ & $0.06$ & $-0.07$ & $0.09$ & $-0.02$ \\
 & $L_{\rm F275W}$ & $0.07$ & $-0.25$ & $-0.05$ & $-0.01$ & $-0.02$ & $-0.09$ \\
 & $f_L$ & $0.05$ & $-0.24$ & $-0.01$ & $-0.02$ & $0.02$ & $-0.01$ \\
 & $\widetilde{\Sigma}_{\rm cl}$ & $-0.02$ & $-0.05$ & $-0.05$ & $-0.04$ & $-0.04$ & $-0.03$ \\
\hline
Common physical scale & $f_N$ & $0.07$ & $-0.11$ & $0.08$ & $-0.04$ & $0.13$ & $-0.05$ \\
 & $L_{\rm F275W}$ & $0.08$ & $-0.29$ & $-0.25$ & $-0.35$ & $-0.04$ & $-0.23$ \\
 & $f_L$ & $0.15$ & $-0.28$ & $-0.14$ & $-0.25$ & $0.04$ & $-0.19$ \\
 & $\widetilde{\Sigma}_{\rm cl}$ & $0.19$ & $0.00$ & $0.05$ & $0.00$ & $0.03$ & $0.03$ \\
\hline
\end{tabular}
\par\smallskip
\begin{minipage}{\textwidth}
\footnotesize
\textit{Notes.} Entries are medians of $\Delta\log\ X=\log\ X_{\rm bar}-\log\ X_{\rm control}$ across the matched sets retained for each selection. Positive offsets indicate larger barred-galaxy measurements. Here $f_N$ and $f_L$ denote regional clump number and F275W luminosity fractions, and $\widetilde{\Sigma}_{\rm cl}=\mathrm{median}_{i}(L_i/A_i)$ within each galaxy's selected region.
Fiducial denotes the baseline analysis. The contrast-threshold test changes only \texttt{min\_delta} to $3\sigma_{\rm sky}$. Physical PSF matching restricts the barred--control PSF-scale difference to less than 10 pc. The common minimum footprint test retains the fiducial matching and applies the largest five-pixel physical area within each matched set to all its members.
\end{minipage}
\end{table*}

We repeated the clump-identification and spatial-profile analysis with \texttt{min\_delta}$=3\sigma_{\rm sky}$, compared with the fiducial value of $1\sigma_{\rm sky}$. All other detection, S/N-selection, and sample-matching criteria were unchanged. This higher contrast threshold reduces the number of retained clumps from 95,486 to 42,526, as fewer low-contrast structures are identified as separate dendrogram leaves.

The bar-interior deficit, recovery toward the bar ends, and bar-aligned concentration remain present (Figure~\ref{fig:min_delta_test}). The regional barred--control offsets are close to their fiducial values (Table~\ref{tab:compact_robustness_medians}). The bar-aligned luminosity-fraction offset also remains at approximately $+0.08$ dex. Thus, the principal spatial signatures are retained despite the substantial change in the number of detected clumps.

\subsection{Sensitivity to physical resolution and minimum clump area}
\label{app:resolution_test}

The fixed value \texttt{min\_npix}$=5$ corresponds to the same angular footprint in every image but not to the same physical area. We therefore examined the sensitivity of the spatial profiles to physical resolution and minimum clump area.

First, we repeated the spatial-profile analysis after requiring the physical F275W PSF scales of each barred galaxy and its controls to agree within 10 pc. This more restrictive matching yields 46 barred galaxies and 53 distinct controls. The resulting bar-interior deficits and bar-aligned number-fraction excess remain close to the fiducial values (Table~\ref{tab:compact_robustness_medians}).

We also applied a common minimum physical footprint within each matched set. We converted the five-pixel detection criterion into a physical area for each galaxy, adopted the largest of these areas within the set, and excluded smaller clumps from the barred galaxy and its controls. Number fractions are recalculated after this cut within $R/R_{\rm eff}\leq2$, while the total galaxy luminosity used in $f_L$ is unchanged. Controls with no retained clumps are omitted.

With the common minimum footprint applied, the bar-interior luminosity and luminosity-fraction offsets remain at $-0.29$ and $-0.28$ dex, while the number-fraction deficit weakens from $-0.20$ to $-0.11$ dex. The bar-aligned number-fraction excess remains at $+0.13$ dex (Table~\ref{tab:compact_robustness_medians}). These signatures persist, although the radial number-fraction deficit is more sensitive to small-clump selection, and residual blending and incompleteness remain possible.

\section{Regional barred--control differences}
\label{app:Regional_differences}
Table~\ref{tab:compact_all_medians} lists the median regional barred--control offsets and two-sided Wilcoxon signed-rank $p$ values for the full matched sample and its subsamples. For each comparison, we tested the paired logarithmic offsets, $\Delta\log X=\log (X_{{\rm bar}}/X_{{\rm control}})$, where the control measurement is the arithmetic mean of the matched-control values in linear units. Only positive barred and averaged-control values enter these comparisons. The $p$ values are not adjusted for multiple comparisons and do not account for dependence between matched sets that share control galaxies. They therefore provide a descriptive assessment of the paired offsets.

\begin{figure*}
    \centering
    \includegraphics[width=\textwidth]{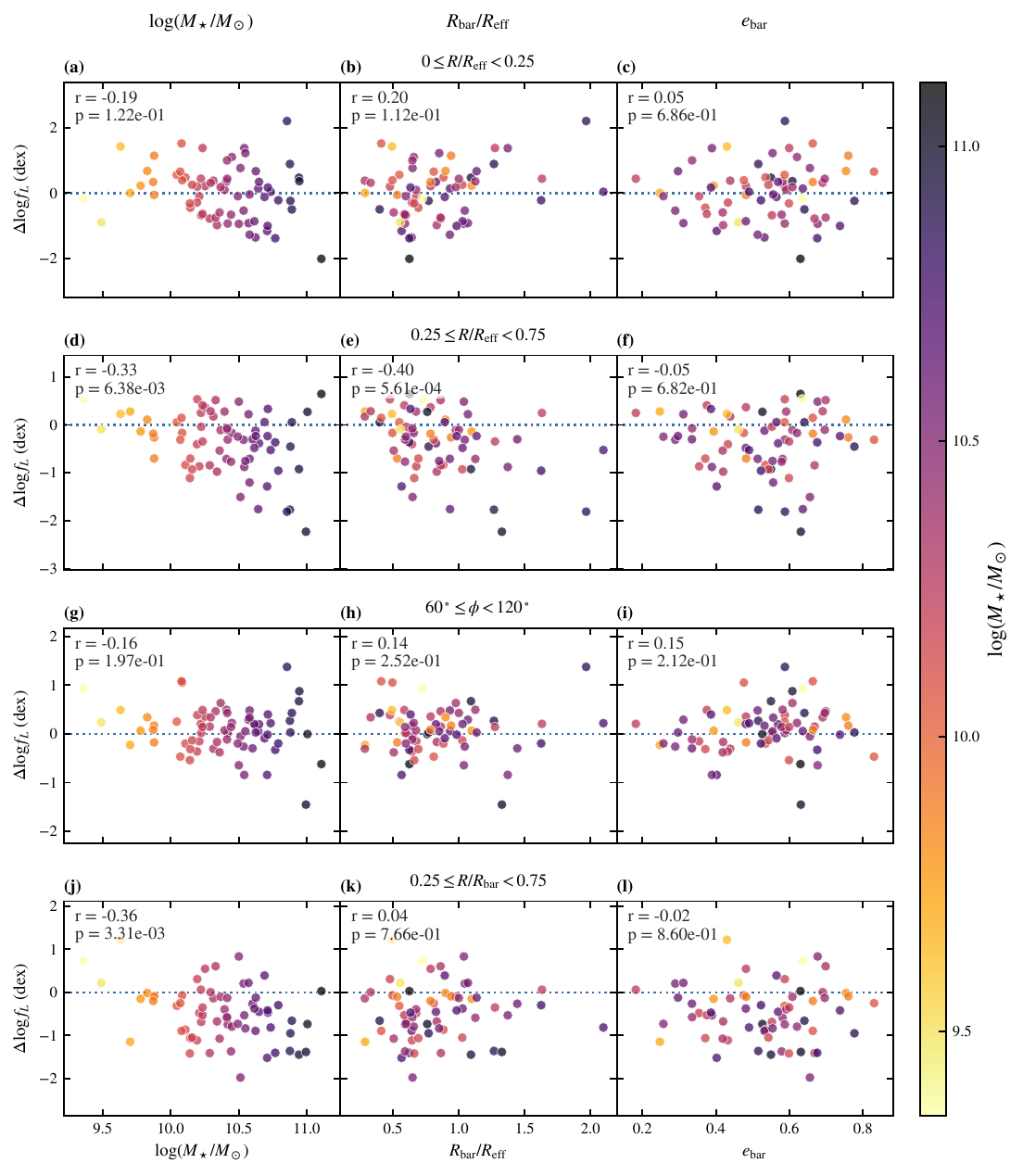}
\caption{Same as Figure~\ref{fig:bar_control_parameter_trends} but for the regional F275W clump luminosity fraction, $f_L$, defined as the summed clump luminosity in each region divided by the total galaxy F275W luminosity. Differences are expressed as $\Delta\log f_L=\log\ f_{L,\rm bar}-\log\ f_{L,\rm control}$, using the same control averaging and plotting conventions.}
\label{fig:bar_control_parameter_trends_luminosity}
\end{figure*}

Figure~\ref{fig:bar_control_parameter_trends_luminosity} shows the galaxy-by-galaxy luminosity-fraction offsets as functions of these properties, complementing the number-fraction comparisons in Figure~\ref{fig:bar_control_parameter_trends}.

\begin{table*}[t]
\caption{Median regional barred--control differences $\Delta\log X$ (dex) with nominal Wilcoxon $p$ values in parentheses.}
\label{tab:compact_all_medians}
\centering
\footnotesize
\setlength{\tabcolsep}{4pt}
\begin{tabular}{lrrrrrr}
\hline\hline
Diagnostic & \multicolumn{3}{c}{Radial range ($R/R_{\rm eff}$)} & \multicolumn{3}{c}{Azimuthal region ($\phi$, deg)} \\
\cline{2-4}\cline{5-7}
 & 0--0.25 & 0.25--0.75 & 0.75--1.25 & 0--60 & 60--120 & 120--180 \\
\hline
\multicolumn{7}{l}{\textit{A. Full sample}}\\
\multicolumn{7}{l}{\textit{All}}\\
$f_N$ & $-0.07\,{\scriptstyle(1.47\times10^{-1})}$ & $-0.20\,{\scriptstyle(1.97\times10^{-6})}$ & $0.06\,{\scriptstyle(1.95\times10^{-1})}$ & $-0.06\,{\scriptstyle(2.35\times10^{-4})}$ & $0.10\,{\scriptstyle(6.07\times10^{-7})}$ & $-0.04\,{\scriptstyle(1.17\times10^{-3})}$ \\
$L_{\rm F275W}$ & $-0.01\,{\scriptstyle(7.08\times10^{-1})}$ & $-0.25\,{\scriptstyle(2.17\times10^{-4})}$ & $-0.10\,{\scriptstyle(5.03\times10^{-2})}$ & $-0.08\,{\scriptstyle(4.26\times10^{-2})}$ & $0.00\,{\scriptstyle(8.31\times10^{-1})}$ & $-0.08\,{\scriptstyle(4.73\times10^{-1})}$ \\
$f_L$ & $0.10\,{\scriptstyle(8.17\times10^{-1})}$ & $-0.26\,{\scriptstyle(4.98\times10^{-5})}$ & $-0.01\,{\scriptstyle(5.18\times10^{-1})}$ & $-0.07\,{\scriptstyle(7.10\times10^{-2})}$ & $0.08\,{\scriptstyle(7.29\times10^{-2})}$ & $-0.09\,{\scriptstyle(2.53\times10^{-1})}$ \\
$\widetilde{\Sigma}_{\rm cl}$ & $0.02\,{\scriptstyle(4.35\times10^{-1})}$ & $-0.03\,{\scriptstyle(4.79\times10^{-1})}$ & $-0.03\,{\scriptstyle(2.09\times10^{-1})}$ & $-0.01\,{\scriptstyle(9.42\times10^{-1})}$ & $-0.03\,{\scriptstyle(4.99\times10^{-1})}$ & $0.00\,{\scriptstyle(9.65\times10^{-1})}$ \\
\hline
\multicolumn{7}{l}{\textit{B. Stellar mass}}\\
\multicolumn{7}{l}{\textit{Low mass}}\\
$f_N$ & $-0.07\,{\scriptstyle(3.07\times10^{-1})}$ & $-0.16\,{\scriptstyle(3.79\times10^{-4})}$ & $0.10\,{\scriptstyle(1.30\times10^{-2})}$ & $-0.04\,{\scriptstyle(8.17\times10^{-2})}$ & $0.07\,{\scriptstyle(9.94\times10^{-4})}$ & $-0.04\,{\scriptstyle(4.05\times10^{-3})}$ \\
$L_{\rm F275W}$ & $-0.01\,{\scriptstyle(7.45\times10^{-1})}$ & $-0.16\,{\scriptstyle(1.06\times10^{-2})}$ & $-0.07\,{\scriptstyle(3.27\times10^{-1})}$ & $-0.24\,{\scriptstyle(2.15\times10^{-2})}$ & $-0.04\,{\scriptstyle(9.42\times10^{-1})}$ & $-0.16\,{\scriptstyle(2.11\times10^{-1})}$ \\
$f_L$ & $0.21\,{\scriptstyle(2.71\times10^{-1})}$ & $-0.15\,{\scriptstyle(3.32\times10^{-2})}$ & $0.08\,{\scriptstyle(1.97\times10^{-1})}$ & $0.02\,{\scriptstyle(4.68\times10^{-1})}$ & $0.14\,{\scriptstyle(5.66\times10^{-2})}$ & $0.01\,{\scriptstyle(9.52\times10^{-1})}$ \\
$\widetilde{\Sigma}_{\rm cl}$ & $0.00\,{\scriptstyle(9.52\times10^{-1})}$ & $-0.07\,{\scriptstyle(5.66\times10^{-2})}$ & $-0.06\,{\scriptstyle(1.15\times10^{-2})}$ & $-0.06\,{\scriptstyle(3.21\times10^{-2})}$ & $-0.05\,{\scriptstyle(6.22\times10^{-2})}$ & $-0.05\,{\scriptstyle(3.81\times10^{-2})}$ \\
\multicolumn{7}{l}{\textit{High mass}}\\
$f_N$ & $0.03\,{\scriptstyle(3.61\times10^{-1})}$ & $-0.32\,{\scriptstyle(5.51\times10^{-4})}$ & $0.00\,{\scriptstyle(4.52\times10^{-1})}$ & $-0.07\,{\scriptstyle(2.09\times10^{-4})}$ & $0.14\,{\scriptstyle(3.05\times10^{-5})}$ & $-0.05\,{\scriptstyle(7.67\times10^{-2})}$ \\
$L_{\rm F275W}$ & $-0.09\,{\scriptstyle(8.97\times10^{-1})}$ & $-0.42\,{\scriptstyle(7.46\times10^{-3})}$ & $-0.21\,{\scriptstyle(8.79\times10^{-2})}$ & $0.09\,{\scriptstyle(5.70\times10^{-1})}$ & $0.13\,{\scriptstyle(1.00\times10^{0})}$ & $0.09\,{\scriptstyle(5.98\times10^{-1})}$ \\
$f_L$ & $-0.12\,{\scriptstyle(3.86\times10^{-1})}$ & $-0.45\,{\scriptstyle(2.17\times10^{-4})}$ & $-0.18\,{\scriptstyle(3.64\times10^{-2})}$ & $-0.12\,{\scriptstyle(7.32\times10^{-2})}$ & $0.04\,{\scriptstyle(5.84\times10^{-1})}$ & $-0.20\,{\scriptstyle(9.19\times10^{-2})}$ \\
$\widetilde{\Sigma}_{\rm cl}$ & $0.09\,{\scriptstyle(2.90\times10^{-1})}$ & $0.00\,{\scriptstyle(2.75\times10^{-1})}$ & $0.01\,{\scriptstyle(4.52\times10^{-1})}$ & $0.07\,{\scriptstyle(2.08\times10^{-2})}$ & $0.03\,{\scriptstyle(2.80\times10^{-1})}$ & $0.07\,{\scriptstyle(7.67\times10^{-2})}$ \\
\hline
\multicolumn{7}{l}{\textit{C. Stellar mass and normalized bar length}}\\
\multicolumn{7}{l}{\textit{Low mass, short bar}}\\
$f_N$ & $-0.05\,{\scriptstyle(5.32\times10^{-1})}$ & $-0.11\,{\scriptstyle(4.35\times10^{-2})}$ & $0.08\,{\scriptstyle(8.42\times10^{-2})}$ & $-0.05\,{\scriptstyle(2.47\times10^{-1})}$ & $0.02\,{\scriptstyle(1.50\times10^{-1})}$ & $-0.02\,{\scriptstyle(2.47\times10^{-1})}$ \\
$L_{\rm F275W}$ & $-0.17\,{\scriptstyle(7.08\times10^{-1})}$ & $0.00\,{\scriptstyle(1.99\times10^{-1})}$ & $-0.05\,{\scriptstyle(4.37\times10^{-1})}$ & $-0.16\,{\scriptstyle(8.42\times10^{-2})}$ & $-0.04\,{\scriptstyle(9.40\times10^{-1})}$ & $-0.16\,{\scriptstyle(2.80\times10^{-1})}$ \\
$f_L$ & $-0.03\,{\scriptstyle(8.42\times10^{-1})}$ & $-0.08\,{\scriptstyle(3.53\times10^{-1})}$ & $0.06\,{\scriptstyle(1.90\times10^{-1})}$ & $0.04\,{\scriptstyle(7.27\times10^{-1})}$ & $0.08\,{\scriptstyle(2.47\times10^{-1})}$ & $-0.01\,{\scriptstyle(8.61\times10^{-1})}$ \\
$\widetilde{\Sigma}_{\rm cl}$ & $-0.06\,{\scriptstyle(3.15\times10^{-1})}$ & $-0.10\,{\scriptstyle(7.51\times10^{-2})}$ & $-0.08\,{\scriptstyle(1.40\times10^{-2})}$ & $-0.09\,{\scriptstyle(1.51\times10^{-2})}$ & $-0.07\,{\scriptstyle(1.11\times10^{-1})}$ & $-0.10\,{\scriptstyle(4.07\times10^{-2})}$ \\
\multicolumn{7}{l}{\textit{Low mass, long bar}}\\
$f_N$ & $-0.07\,{\scriptstyle(5.02\times10^{-1})}$ & $-0.31\,{\scriptstyle(3.05\times10^{-3})}$ & $0.13\,{\scriptstyle(7.85\times10^{-2})}$ & $-0.04\,{\scriptstyle(2.17\times10^{-1})}$ & $0.11\,{\scriptstyle(1.71\times10^{-3})}$ & $-0.08\,{\scriptstyle(1.22\times10^{-3})}$ \\
$L_{\rm F275W}$ & $0.20\,{\scriptstyle(3.91\times10^{-1})}$ & $-0.47\,{\scriptstyle(2.45\times10^{-2})}$ & $-0.07\,{\scriptstyle(5.42\times10^{-1})}$ & $-0.24\,{\scriptstyle(1.35\times10^{-1})}$ & $-0.03\,{\scriptstyle(8.08\times10^{-1})}$ & $-0.15\,{\scriptstyle(5.02\times10^{-1})}$ \\
$f_L$ & $0.40\,{\scriptstyle(2.41\times10^{-1})}$ & $-0.29\,{\scriptstyle(2.45\times10^{-2})}$ & $0.12\,{\scriptstyle(7.15\times10^{-1})}$ & $-0.10\,{\scriptstyle(6.26\times10^{-1})}$ & $0.16\,{\scriptstyle(1.04\times10^{-1})}$ & $0.03\,{\scriptstyle(8.55\times10^{-1})}$ \\
$\widetilde{\Sigma}_{\rm cl}$ & $0.02\,{\scriptstyle(2.17\times10^{-1})}$ & $-0.05\,{\scriptstyle(3.91\times10^{-1})}$ & $-0.04\,{\scriptstyle(3.91\times10^{-1})}$ & $-0.01\,{\scriptstyle(1.00\times10^{0})}$ & $-0.05\,{\scriptstyle(3.91\times10^{-1})}$ & $-0.02\,{\scriptstyle(6.26\times10^{-1})}$ \\
\multicolumn{7}{l}{\textit{High mass, short bar}}\\
$f_N$ & $-0.68\,{\scriptstyle(7.81\times10^{-2})}$ & $0.01\,{\scriptstyle(5.70\times10^{-1})}$ & $0.06\,{\scriptstyle(9.10\times10^{-1})}$ & $-0.09\,{\scriptstyle(2.73\times10^{-2})}$ & $0.12\,{\scriptstyle(1.95\times10^{-2})}$ & $-0.01\,{\scriptstyle(7.34\times10^{-1})}$ \\
$L_{\rm F275W}$ & $-0.48\,{\scriptstyle(5.47\times10^{-1})}$ & $0.06\,{\scriptstyle(9.10\times10^{-1})}$ & $-0.34\,{\scriptstyle(5.70\times10^{-1})}$ & $-0.24\,{\scriptstyle(8.20\times10^{-1})}$ & $0.33\,{\scriptstyle(8.20\times10^{-1})}$ & $0.06\,{\scriptstyle(5.70\times10^{-1})}$ \\
$f_L$ & $-0.83\,{\scriptstyle(7.81\times10^{-2})}$ & $-0.12\,{\scriptstyle(4.26\times10^{-1})}$ & $-0.39\,{\scriptstyle(9.77\times10^{-2})}$ & $-0.28\,{\scriptstyle(9.77\times10^{-2})}$ & $0.00\,{\scriptstyle(7.34\times10^{-1})}$ & $-0.20\,{\scriptstyle(5.47\times10^{-2})}$ \\
$\widetilde{\Sigma}_{\rm cl}$ & $-0.10\,{\scriptstyle(4.61\times10^{-1})}$ & $-0.06\,{\scriptstyle(8.20\times10^{-1})}$ & $-0.02\,{\scriptstyle(7.34\times10^{-1})}$ & $-0.04\,{\scriptstyle(5.70\times10^{-1})}$ & $-0.04\,{\scriptstyle(7.34\times10^{-1})}$ & $-0.06\,{\scriptstyle(7.34\times10^{-1})}$ \\
\multicolumn{7}{l}{\textit{High mass, long bar}}\\
$f_N$ & $0.12\,{\scriptstyle(7.68\times10^{-1})}$ & $-0.41\,{\scriptstyle(3.95\times10^{-4})}$ & $-0.05\,{\scriptstyle(3.20\times10^{-1})}$ & $-0.06\,{\scriptstyle(2.86\times10^{-3})}$ & $0.15\,{\scriptstyle(8.52\times10^{-4})}$ & $-0.05\,{\scriptstyle(6.46\times10^{-2})}$ \\
$L_{\rm F275W}$ & $-0.09\,{\scriptstyle(3.12\times10^{-1})}$ & $-0.59\,{\scriptstyle(8.51\times10^{-4})}$ & $-0.16\,{\scriptstyle(1.37\times10^{-1})}$ & $0.12\,{\scriptstyle(6.58\times10^{-1})}$ & $0.05\,{\scriptstyle(9.19\times10^{-1})}$ & $0.17\,{\scriptstyle(7.85\times10^{-1})}$ \\
$f_L$ & $0.04\,{\scriptstyle(8.29\times10^{-1})}$ & $-0.62\,{\scriptstyle(1.34\times10^{-4})}$ & $-0.15\,{\scriptstyle(1.37\times10^{-1})}$ & $-0.08\,{\scriptstyle(3.04\times10^{-1})}$ & $0.06\,{\scriptstyle(3.93\times10^{-1})}$ & $-0.20\,{\scriptstyle(3.93\times10^{-1})}$ \\
$\widetilde{\Sigma}_{\rm cl}$ & $0.13\,{\scriptstyle(1.23\times10^{-1})}$ & $0.04\,{\scriptstyle(7.59\times10^{-2})}$ & $0.03\,{\scriptstyle(1.91\times10^{-1})}$ & $0.12\,{\scriptstyle(1.37\times10^{-3})}$ & $0.07\,{\scriptstyle(8.22\times10^{-2})}$ & $0.08\,{\scriptstyle(9.02\times10^{-3})}$ \\
\hline
\multicolumn{7}{l}{\textit{D. Stellar mass and bar ellipticity}}\\
\multicolumn{7}{l}{\textit{Low mass, low $e_{\rm bar}$}}\\
$f_N$ & $-0.04\,{\scriptstyle(7.84\times10^{-1})}$ & $-0.13\,{\scriptstyle(2.15\times10^{-2})}$ & $0.18\,{\scriptstyle(4.00\times10^{-2})}$ & $0.02\,{\scriptstyle(9.27\times10^{-1})}$ & $0.04\,{\scriptstyle(1.33\times10^{-1})}$ & $-0.02\,{\scriptstyle(5.32\times10^{-2})}$ \\
$L_{\rm F275W}$ & $-0.24\,{\scriptstyle(4.52\times10^{-1})}$ & $-0.30\,{\scriptstyle(1.53\times10^{-2})}$ & $-0.09\,{\scriptstyle(1.65\times10^{-1})}$ & $-0.43\,{\scriptstyle(1.92\times10^{-2})}$ & $-0.17\,{\scriptstyle(1.14\times10^{-1})}$ & $-0.28\,{\scriptstyle(8.97\times10^{-2})}$ \\
$f_L$ & $-0.03\,{\scriptstyle(8.98\times10^{-1})}$ & $-0.15\,{\scriptstyle(6.96\times10^{-2})}$ & $0.00\,{\scriptstyle(8.12\times10^{-1})}$ & $-0.23\,{\scriptstyle(2.94\times10^{-1})}$ & $-0.14\,{\scriptstyle(7.84\times10^{-1})}$ & $-0.10\,{\scriptstyle(3.88\times10^{-1})}$ \\
$\widetilde{\Sigma}_{\rm cl}$ & $-0.07\,{\scriptstyle(2.77\times10^{-1})}$ & $-0.11\,{\scriptstyle(1.07\times10^{-2})}$ & $-0.10\,{\scriptstyle(9.44\times10^{-3})}$ & $-0.11\,{\scriptstyle(1.53\times10^{-2})}$ & $-0.13\,{\scriptstyle(1.72\times10^{-2})}$ & $-0.13\,{\scriptstyle(2.40\times10^{-2})}$ \\
\multicolumn{7}{l}{\textit{Low mass, high $e_{\rm bar}$}}\\
$f_N$ & $-0.15\,{\scriptstyle(1.14\times10^{-1})}$ & $-0.22\,{\scriptstyle(9.44\times10^{-3})}$ & $0.06\,{\scriptstyle(1.89\times10^{-1})}$ & $-0.08\,{\scriptstyle(9.44\times10^{-3})}$ & $0.11\,{\scriptstyle(2.71\times10^{-3})}$ & $-0.06\,{\scriptstyle(2.66\times10^{-2})}$ \\
$L_{\rm F275W}$ & $0.09\,{\scriptstyle(2.77\times10^{-1})}$ & $0.00\,{\scriptstyle(3.30\times10^{-1})}$ & $-0.03\,{\scriptstyle(9.56\times10^{-1})}$ & $0.09\,{\scriptstyle(3.88\times10^{-1})}$ & $0.17\,{\scriptstyle(8.26\times10^{-2})}$ & $-0.12\,{\scriptstyle(9.56\times10^{-1})}$ \\
$f_L$ & $0.33\,{\scriptstyle(1.43\times10^{-1})}$ & $-0.14\,{\scriptstyle(2.61\times10^{-1})}$ & $0.15\,{\scriptstyle(1.54\times10^{-1})}$ & $0.05\,{\scriptstyle(8.41\times10^{-1})}$ & $0.24\,{\scriptstyle(4.86\times10^{-3})}$ & $0.14\,{\scriptstyle(3.49\times10^{-1})}$ \\
$\widetilde{\Sigma}_{\rm cl}$ & $0.02\,{\scriptstyle(3.30\times10^{-1})}$ & $-0.01\,{\scriptstyle(1.00\times10^{0})}$ & $-0.04\,{\scriptstyle(3.88\times10^{-1})}$ & $0.00\,{\scriptstyle(8.69\times10^{-1})}$ & $-0.02\,{\scriptstyle(9.27\times10^{-1})}$ & $-0.01\,{\scriptstyle(6.48\times10^{-1})}$ \\
\multicolumn{7}{l}{\textit{High mass, low $e_{\rm bar}$}}\\
$f_N$ & $0.03\,{\scriptstyle(8.90\times10^{-1})}$ & $-0.35\,{\scriptstyle(1.50\times10^{-2})}$ & $-0.10\,{\scriptstyle(1.74\times10^{-1})}$ & $-0.07\,{\scriptstyle(3.16\times10^{-3})}$ & $0.16\,{\scriptstyle(2.14\times10^{-4})}$ & $-0.06\,{\scriptstyle(5.69\times10^{-2})}$ \\
$L_{\rm F275W}$ & $-0.19\,{\scriptstyle(8.90\times10^{-1})}$ & $-0.54\,{\scriptstyle(6.65\times10^{-3})}$ & $-0.21\,{\scriptstyle(4.64\times10^{-3})}$ & $-0.12\,{\scriptstyle(2.07\times10^{-1})}$ & $0.11\,{\scriptstyle(7.12\times10^{-1})}$ & $-0.11\,{\scriptstyle(4.59\times10^{-1})}$ \\
$f_L$ & $-0.04\,{\scriptstyle(6.79\times10^{-1})}$ & $-0.52\,{\scriptstyle(1.34\times10^{-3})}$ & $-0.26\,{\scriptstyle(1.74\times10^{-2})}$ & $-0.21\,{\scriptstyle(6.38\times10^{-2})}$ & $0.17\,{\scriptstyle(4.04\times10^{-1})}$ & $-0.20\,{\scriptstyle(5.05\times10^{-2})}$ \\
$\widetilde{\Sigma}_{\rm cl}$ & $0.09\,{\scriptstyle(1.69\times10^{-1})}$ & $0.09\,{\scriptstyle(3.05\times10^{-2})}$ & $0.01\,{\scriptstyle(4.31\times10^{-1})}$ & $0.10\,{\scriptstyle(1.50\times10^{-2})}$ & $0.07\,{\scriptstyle(1.20\times10^{-1})}$ & $0.08\,{\scriptstyle(2.02\times10^{-2})}$ \\
\multicolumn{7}{l}{\textit{High mass, high $e_{\rm bar}$}}\\
$f_N$ & $-0.17\,{\scriptstyle(2.66\times10^{-1})}$ & $-0.31\,{\scriptstyle(2.10\times10^{-2})}$ & $0.07\,{\scriptstyle(8.39\times10^{-1})}$ & $-0.07\,{\scriptstyle(3.98\times10^{-2})}$ & $0.09\,{\scriptstyle(6.81\times10^{-2})}$ & $-0.02\,{\scriptstyle(7.87\times10^{-1})}$ \\
$L_{\rm F275W}$ & $0.18\,{\scriptstyle(7.91\times10^{-1})}$ & $-0.16\,{\scriptstyle(3.39\times10^{-1})}$ & $0.15\,{\scriptstyle(9.46\times10^{-1})}$ & $0.15\,{\scriptstyle(6.85\times10^{-1})}$ & $0.33\,{\scriptstyle(5.42\times10^{-1})}$ & $0.26\,{\scriptstyle(1.91\times10^{-1})}$ \\
$f_L$ & $-0.18\,{\scriptstyle(5.19\times10^{-1})}$ & $-0.37\,{\scriptstyle(1.10\times10^{-1})}$ & $-0.06\,{\scriptstyle(4.97\times10^{-1})}$ & $0.00\,{\scriptstyle(5.88\times10^{-1})}$ & $-0.02\,{\scriptstyle(8.93\times10^{-1})}$ & $-0.17\,{\scriptstyle(7.87\times10^{-1})}$ \\
$\widetilde{\Sigma}_{\rm cl}$ & $0.04\,{\scriptstyle(9.10\times10^{-1})}$ & $-0.08\,{\scriptstyle(5.19\times10^{-1})}$ & $0.02\,{\scriptstyle(9.46\times10^{-1})}$ & $-0.01\,{\scriptstyle(5.42\times10^{-1})}$ & $-0.04\,{\scriptstyle(8.39\times10^{-1})}$ & $-0.02\,{\scriptstyle(9.46\times10^{-1})}$ \\
\hline
\end{tabular}
\par\smallskip
\begin{minipage}{\textwidth}
\footnotesize
\textit{Notes.} Each entry gives the median paired offset $\Delta\log X$in dex followed by the nominal two-sided Wilcoxon signed-rank $p$ value in parentheses. The paired-offset definition, diagnostic definitions, and control averaging follow Table~\ref{tab:compact_robustness_medians}, using the fiducial sample of 70 barred galaxies and 95 distinct controls. The $p$ values are not corrected for multiple comparisons and do not account for dependence from shared controls. The fourth diagnostic, $\widetilde{\Sigma}_{\rm cl}$, is the within-galaxy median clump luminosity surface density. Low/high mass uses $\log(M_\star/M_\odot)\leq10.5$/$>10.5$, short/long bars use $R_{\rm bar}/R_{\rm eff}<0.8$/$\geq0.8$, and low/high ellipticity uses $e_{\rm bar}<0.6$/$\geq0.6$.
\end{minipage}
\end{table*}

\end{appendix}
%
%
\end{document}